\documentclass[twocolumn,showpacs,%
  nofootinbib,aps,superscriptaddress,%
  eqsecnum,prd,notitlepage,showkeys,10pt]{revtex4-1}

\usepackage{amssymb}
\usepackage{xcolor}
\usepackage{amsmath}
\usepackage{graphicx, wrapfig, lipsum, caption}
\usepackage{subfigure}
\usepackage{dcolumn}
\usepackage{hyperref}
\usepackage{blindtext}
\usepackage{setspace}
\usepackage{booktabs}
\usepackage{romannum}

\usepackage{soul}

\begin{document}

\title{Benchmarking Pedestrian Dynamics Models for Common Scenarios: An Evaluation of Force-Based Models}

\author{Kanika Jain$^1$, Shankar Prawesh$^{2*}$, Indranil Saha Dalal$^{1*}$, Anurag Tripathi$^*$}

\affiliation{Department of Chemical Engineering, Indian Institute of Technology Kanpur, India\\$^2$Department of Management Sciences, Indian Institute of Technology Kanpur, India}

\begin{abstract}
\textbf{Abstract:} Extensive research in pedestrian dynamics has primarily focused on crowded conditions and associated phenomena, such as lane formation, evacuation, etc. Several force-based models have been developed to predict the behavior in these situations. In contrast, there is a notable gap in terms of investigations of the moderate-to-low density situations. These scenarios are extremely commonplace across the world, including the highly populated nations like India. Additionally, the details of force-based models are expected to show significant effects at these densities, whereas the crowded, nearly packed, conditions may be expected to be governed largely by contact forces. In this study, we address this gap and comprehensively evaluate the performance of different force-based models in some common scenarios. Towards this, we perform controlled experiments in four situations: avoiding a stationary obstacle, position-swapping by walking toward each other, overtaking to reach a common goal, and navigating through a maze of obstacles. The performance evaluation consists of two stages and six evaluating metrics - successful trajectories, overlapping proportion, oscillation strength, path smoothness, speed deviation, and travel time. Firstly, models must meet an eligibility criterion of at least 80\% successful trajectories and secondly, the models are scored based on the cutoff values established from the experimental data. We evaluated five force-based models where the best one scored 57.14\%. Thus, our findings reveal significant shortcomings in the ability of these models to yield accurate predictions of pedestrian dynamics in these common situations.
\end{abstract}

\maketitle

\begin{figure*} [t]
    \includegraphics[width= 1\linewidth]{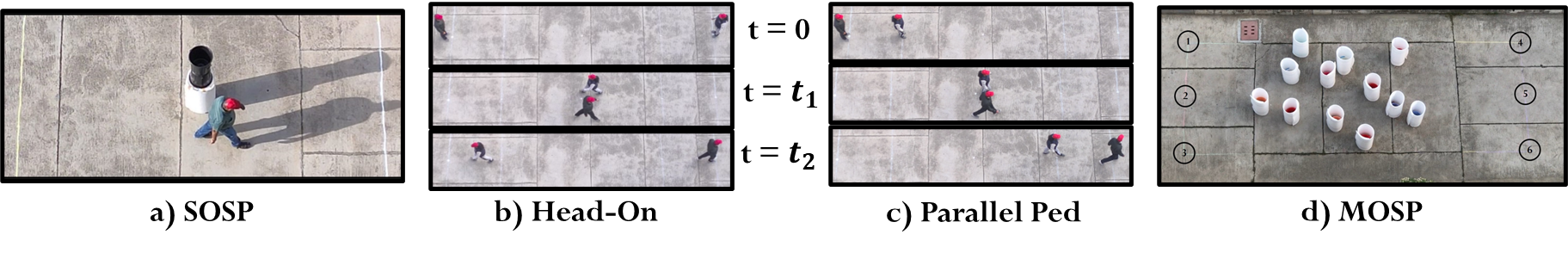}
    \caption{{\footnotesize Controlled experiments illustrating four real-life scenarios. (a) Single Obstacle Single Pedestrian (SOSP): a volunteer passing through a non-living, human-sized stationary obstacle to reach their goal. (b) Head-On: two volunteers approaching from opposite directions, swapping their initial positions. (c) Parallel Ped: a faster volunteer overtaking a slower one while walking in the same direction aiming to a common goal. Snapshots at three different time intervals display the initial condition ($t = 0$), midway condition ($t = t_1$), and final condition ($t = t_2$) for head-on and parallel ped scenarios. d) Multiple Obstacles Single Pedestrian (MOSP): a volunteer navigating through a randomly placed maze of obstacles to reach their goal.}} \label{fig:experiments}
\end{figure*}

\section{Introduction} \label{sec:Intro}

The dynamics of pedestrians has been studied by various models proposed over past few decades by different researchers. These include the perspectives of the motion of individual pedestrians in any given scenario as well as the evolution of the state of groups of pedestrians. The types of models also vary widely and range from force-based, using classical laws of motion \cite{karamouzas2014universal, helbing1995social, chraibi2010generalized}, to heuristic, rule-based approaches \cite{moussaid2011simple}. These studies have been accompanied by appropriate experimental ones as well, across various nations and situations \cite{helbing2007dynamics, haghani2017following, appert2018experimental, moussaid2009experimental}. The central problem, one of being able to accurately model pedestrian motion or the state of a group of pedestrians (or a crowd, which usually refers to a large number of pedestrians at high density), is of enormous significance. This is expected to enable the fundamental physics-based design of public infrastructure, where the aspects of safety and mitigation of stampede-like disasters. Such successful models will have a positive impact on public space designs across the world, more significantly in countries like India, with high population densities.

\begin{table}[h]
    \centering

    \caption{{\footnotesize Details of experiments (mimicking real-life scenarios) performed with the number of runs for each case.}}
    \label{tab:data}

    \scriptsize
    \def\arraystretch{2}
    
    \begin{tabular}{|c|c|}
    \hline
        \textbf{Experiment} & \textbf{$\#$ runs} \\\hline
        SOSP & 58\\\hline
        Head-On & 25\\\hline
        Parallel-Ped & 27\\\hline
        MOSP & \\
        Case A: 3.74\% & 239\\
        Case B: 6.54\% & 188\\
        Case C: 11.22\% & 184\\
        Case D: 14.96\% & 276\\\hline
    \end{tabular}
\end{table}

As discussed, various models have been proposed by earlier researchers, showing great promise to mimic different pedestrian scenarios. These models must be systematically evaluated to determine which one is most appropriate for a given application, based on how well it captures the relevant properties of the scenario. We anticipate that the evaluation of the effectiveness of these models will involve various situations - ranging from those involving the motion of many pedestrians to a few, with or without other obstacles (stationary, may or may not be involving other human beings) around them. The former, or crowded scenarios, include lane formation \cite{feliciani2016empirical, schadschneider2011empirical, kramer2021social}, bottleneck \cite{schadschneider2011empirical, helbing2006analytical}, evacuation \cite{von2017empirical, shiwakoti2019review, heliovaara2012pedestrian, lui2012constructal, lui2013constructal}, and crowd intersection \cite{gorrini2013empirical, shiwakoti2015empirical, lian2015experimental}, extensively documented in the literature. Evaluation metrics for these scenarios include fundamental diagrams \cite{seyfried2005fundamental, seyfried2007fundamental, seyfried2010enhanced}, density measure \cite{lerner2009data}, velocity fields \cite{liddle2022microscopic}, etc. There are empirical studies on different crowded scenarios in the literature \cite{haghani2018crowd, haghani2020empirical}. These studies provide benchmarks for assessing model performance in crowded conditions. However, note that all these situations involve a high density crowd, at closely packed conditions. Under such conditions, the close contact interactions are expected to govern pedestrian motion rather than the fundamental interactions between them, which might be active when they have some gaps around each other to move and manoeuvre. Thus, almost all reasonable force-based models (and maybe even rule based ones) may be expected to converge to similar answers for such highly packed scenarios. Interesting differences may be anticipated in moderate-to-low density situations, where the details of the interaction between pedestrians at some distance (i.e. not in close contact) are likely to influence the motion of individual pedestrians. Interestingly, in sharp contrast to crowded scenarios, much fewer studies focus on moderate-to-low density scenarios. Typically, such studies focus on the measurement of specific features like personal space \cite{shan2014critical, parisi2016experimental}, gait features \cite{cao2018stepping, renaudin2012step, racic2009experimental}, etc., rather than the comparison of motion of pedestrians in situations from experiments and models. Thus, since such situations are not clearly established, the aspect of model evaluation with varying details of forces (and rules) remains incomplete. Hence, there exists a critical gap in terms of establishing benchmarks for evaluating pedestrian dynamics models, which we would address in this article. Note, a gap also exists for such comparisons involving Indian pedestrians, as most of the earlier investigations were performed in western or eastern nations. In this context, we note that such moderate-to-low density conditions are extremely common in various public spaces at all times, across India and other nations with large population densities. Our investigations here also extends to high density situations involving one moving pedestrian among many stationary obstacles (may or may not be humans). In this context, in a previous study, we have shown that the behavior of a single pedestrian around a non-living obstacle is similar to that around a similar-sized human being \cite{jainAutomatic}. Ideally, we expect all the models to perform reasonably well at both high, moderate and low densities, but our study here proves the contrary (as shown later in the article).

From the perspective discussed above, this study addresses this gap for various force-based models in the literature. We focus on common situations at the moderate-to-low density regimes, where the details of interactions will govern the pedestrian motion. In the process, we have two broad objectives. Firstly, the benchmarking criteria for such situations are established, such that any future models can be evaluated similarly. Secondly, various force-based models present in the literature are evaluated within this criteria. Towards this, our study performs controlled experiments across four commonplace real-life scenarios (Figure \ref{fig:experiments}), and respective data collection is given in Table \ref{tab:data}. As explained later, these situations include one or two pedestrians, in the absence or presence of other stationary obstacles. These obstacles may represent other human beings (stationary) or inanimate objects (like pillars or similar objects). The number of obstacles is varied to traverse various regimes of density. Bottinelli and Silverburg observe instances where high-density crowds behave similarly to soft solids \cite{bottinelli2019dense}. Here, the term “density” is used in the broader sense of the total number of objects within the experimental area, including the pedestrians. For evaluation across models, we use six evaluating metrics - successful trajectories, overlapping proportion, oscillation strength, path smoothness, speed deviation, and travel time, as explained in the article. There are many force-based models present in the literature \cite{karamouzas2014universal, helbing2000simulating, helbing1995social, chraibi2009quantitative, chraibi2010generalized, hu2023anticipation}. We evaluate five that are frequently used for applications - Universal Power Law (UPL) \cite{karamouzas2014universal}, Social Force Model - circular (SFMc) \cite{helbing2000simulating}, Social Force Model - elliptical (SFMe) \cite{helbing1995social}, Centrifugal Force Model - circular (CFMc) \cite{chraibi2009quantitative}, and Centrifugal Force Model - elliptical (CFMe) \cite{chraibi2010generalized}. These models were tested against the selected situations and evaluated using a scoring system based on a two-stage evaluation process, as explained later. In the first stage, models are assessed against specific eligibility criteria to determine their suitability. In the second stage, a finer comparison of the models is performed based on cutoff values established from experimental data. As highlighted later, our findings reveal significant deficiencies in the ability of force-based models to predict experimental pedestrian behavior. To regenerate the experimental scenarios and calculate evaluating metrics for scoring, initial position, final position, and desired speed for each volunteer and maximum travel time in each scenario are provided \url{https://github.com/kanika201293/Pedestrian-Experimental-Data}. 

\section{Methodology} \label{sec:methodology}

\begin{figure}
    \includegraphics[width= 1\linewidth]{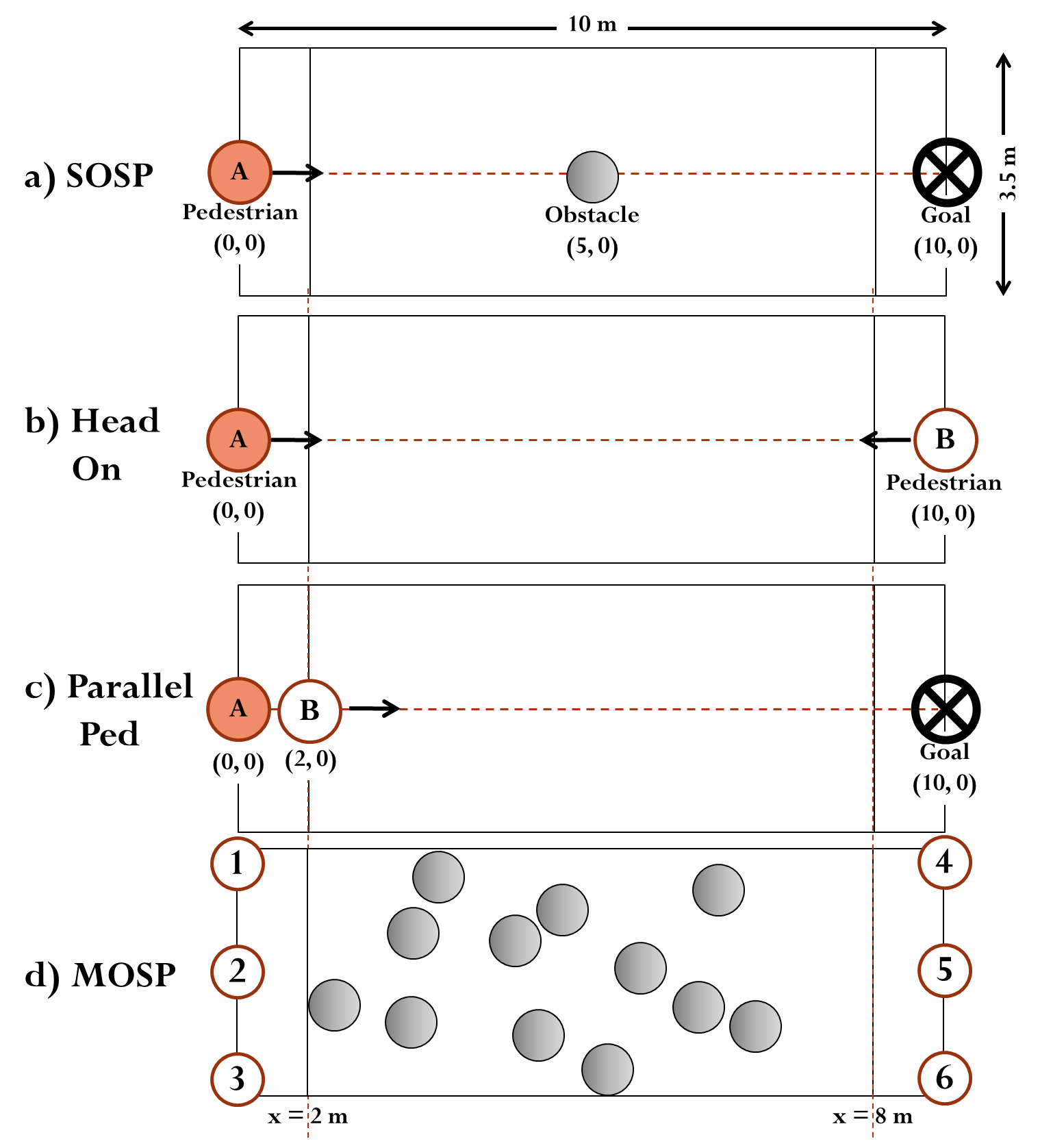}
    \caption{{\footnotesize Schematic of all the experimental setups within a boundary defined as $x = 0\;m$ to $x = 10\;m$ and $y = -1.75\;m$ to $y = 1.75\;m$. The measurement region spans from $x = 2\;m$ to $x = 8\;m$. For future reference, volunteer/simulated pedestrian A is represented by a filled circle, and an empty circle represents volunteer/simulated pedestrian B.}} \label{fig:experimentalSetups}
\end{figure}

\subsection{Experimental Setup}

In this study, we conducted controlled experiments for four different low to moderate density real-life-based scenarios. Note, all these scenarios involve 1-2 pedestrians, but some also have a high density of stationary obstacles. However, all of these represent commonly occurring everyday situations. Note, from the point of view of the moving pedestrian, an obstacle is equivalent to a stationary similar-sized human, as shown in our earlier study \cite{jainAutomatic}. We selected obstacles that have similar diameter as a human being. Table \ref{tab:data} provides details of the experiments and data collection procedure. The experimental boundary was defined as a rectangular area with $x = 0\;m$ to $x = 10\;m$ and $y = -1.75\;m$ to $y = 1.75\;m$, with the measurement region spanning from $x = 2\;m$ to $x = 8\;m$. The experiments conducted are as follows:\\\\
\textbf{1.\;Single Obstacle Single Pedestrian (SOSP): }In this scenario, a single volunteer must reach a specified goal by avoiding a non-living, human-sized stationary obstacle placed in the middle of the path. Figure \ref{fig:experimentalSetups}(a) shows the schematic of the setup.\\\\
\textbf{2.\;Head-On: }In the head-on scenario, two volunteers, denoted as A and B, walk towards each other at their desired speeds with the objective of exchanging their initial positions to reach their respective goals. A schematic illustrating this scenario is shown in Figure \ref{fig:experimentalSetups}(b). Thus, the initial position of A is the goal of B and vice-versa.\\\\
\textbf{3.\;Parallel-Ped: }In the parallel-ped scenario, two volunteers, denoted as A and B, share a common destination, as illustrated in Figure \ref{fig:experimentalSetups}(c). Volunteer A is initially positioned at a separation of 2 meters, behind volunteer B and has a tendency to walk at a faster pace than volunteer B. Consequently, it is expected that volunteer A will surpass B and both of them will walk towards the common goal.\\\\
\textbf{4. Multiple Obstacles Single Pedestrian (MOSP): }In this case, a single pedestrian passes through a maze of randomly placed obstacles to reach the goal. A series of experiments are performed by varying the area fraction occupied by the obstacles. Figure \ref{fig:experimentalSetups}(d) gives a schematic of randomly placed obstacles covering 11.22\% of the area. The experiment is repeated for four different area fractions: 3.74\%, 6.54\%, 11.22\%, and 14.96\%. The initial and final positions of a volunteer were selected randomly between position numbers 1 to 6, marked by circles (see Figure \ref{fig:experimentalSetups}(d)).

\subsection{Evaluating metrics}
\label{sec:evaluatingmetrics}

The following are the metrics used in this study to evaluate the performance of any model, for any given scenario.\\\\
\textbf{Successful Trajectories: }A trajectory is defined as \textit{unsuccessful} when it fulfills any of the following conditions in the simulation.

\begin{itemize}
    \item The time taken by the pedestrian to reach the goal exceeds twice the maximum travel time in the corresponding experimental setup, i.e., the pedestrian is stuck in between the path. Note that no stuck condition is observed during experiments.
    \item At any instant, the maximum overlap between a pedestrian and an obstacle or another pedestrian exceeds fifty percent.
    \item At any instant, more than fifty percent of the pedestrian's body extends outside the experimental setup boundary.
\end{itemize}

The number of successful trajectories for a particular scenario is determined by subtracting the number of unsuccessful trajectories from the total number of trajectories.\\\\
\textbf{Overlapping Proportion:} 
``Overlapping'' between two pedestrians or between a pedestrian and an obstacle occurs when their geometrical form overlaps. In this study, all the pedestrians and obstacles are modeled as circles. A measure of the extent of overlap is defined earlier by Chraibi et. al. \cite{chraibi2010generalized}, as follows:

\begin{align}
   o^{(v)} &= \dfrac{1}{nov} \sum_{t=0}^{t=t_{end}} \sum_{i=1}^{i=N} \sum_{j>i}^{j=N} o{ij} \label{eq:overlap}\\    
    o_{ij} &= \dfrac{A_{ij}}{min(A_i,A_j)} \le 1 \notag
\end{align}

Here, $N$ represents the total count of simulated pedestrians, $A_i$ and $A_j$ denote the areas of the circular pedestrians $i$ and $j$, respectively, and the variable $A_{ij}$ represents the overlapping area between the two. In (\ref{eq:overlap}), the expression is normalized by the parameter $nov$, representing the number of simulation steps where overlapping appears. $nov$ is $0$ if there is no overlapping during the whole simulation. In order to evaluate the models thoroughly, the maximum overlap is also calculated in some cases.\\\\
\textbf{Oscillation Strength:} Backward movements of a pedestrian (opposite to the direction of the goal) during a simulation are called oscillations. Such movements are unrealistic in nature and need to be addressed in a model. To quantify the oscillations, we used the following measure introduced by Chraibi et al. \cite{chraibi2010generalized}:

\begin{align}
    o^{(s)} &= \dfrac{1}{nos} \sum_{t=0}^{t=t_{end}} \sum_{i=1}^{i=N} S_i \label{eq:oscillate}\\
    S_i &= \dfrac{1}{2} (-s_i + |s_i|) \notag \\
    s_i &= \dfrac{\overrightarrow {v_i} \cdot \overrightarrow {v_i}^0}{(v_i^0)^2} \notag
\end{align}

Here, $N$ represents the total count of simulated pedestrians, $\overrightarrow {v_i}$ is the current velocity, and $\overrightarrow {v_i^0}$ is the desired velocity of pedestrian $i$. In (\ref{eq:oscillate}), the expression is normalized by the parameter $nos$, representing the number of simulation steps where oscillation appears; $nos$ is set to $0$ if there is no oscillation during the whole simulation. Note that the desired velocity is the desired speed of the pedestrian towards the goal.\\\\
\textbf{Path Smoothness: }This metric measures the smoothness of a path by identifying any sudden changes in path direction. It is calculated by measuring the maximum absolute change in the path angle which is calculated at each time instant. Let $\overrightarrow{P_t}$ be the position vector of the pedestrian at time instant $t$. The displacement vector $\overrightarrow{V_t}$ is defined as the change between position vectors $\overrightarrow{P_t}$ and $\overrightarrow{P_{t+1}}$. The angular deviation of path,  $\theta_t$ is the angle between the displacement vectors $\overrightarrow{V_t}$ and $\overrightarrow{V_{t+1}}$. Path smoothness, $\theta_i$, for pedestrian $i$, is then the maximum of the absolute value of angles $\theta_t$. The lower value of $\theta_i$ denotes higher path smoothness for the pedestrian. The quantitative calculation of path smoothness is as follows:

\begin{align}
    PS &= \dfrac{1}{N} \sum_{i=1}^{i=N}\theta_i\label{eq:pathSmoothness}\\
    \overrightarrow{P_t} &: \text{position vector at }t\;\text{time instant}\notag \\
    \overrightarrow{V_t} &: \text{displacement vector } (\overrightarrow{P_{t+1}}-\overrightarrow{P_t})\notag \\ 
    \theta_t &: \text{angle between vectors } (\overrightarrow{V_{t+1}}-\overrightarrow{V_t})\notag \\
    \theta_i &: max\{\theta_t\} \notag
\end{align}

Here, $N$ refers to the total count of simulated pedestrians where pedestrians are denoted by $i$.\\\\
\textbf{Speed Deviation: }Speed deviation is defined as the average of the absolute change in speed relative to the desired speed at each time instant (t).

\begin{align}
    SD &= \dfrac{1}{N} \sum_{i=1}^{i=N} \dfrac{1}{nsd} \sum_{t=0}^{t=t_{end}}\dfrac{|v_i^0 - v_i(t)|}{v_i^0}
    \label{eq:speedDeviation}
\end{align}

Here, $v_i^0$ represents the desired speed, and $v_i(t)$ represents the speed at time instant $t$, respectively for pedestrian $i$. $nsd$ refers to the number of simulation steps and $N$ refers to the total count of simulated pedestrians.\\\\
\textbf{Travel Time: }This metric represents the total time taken by a pedestrian to traverse the measurement area. During simulation, a trajectory is deemed unsuccessful if the time taken by the simulator exceeds twice the maximum travel time recorded during the experiment.\\\\
Note the following for our calculations in this study:\\ 1) Single values of evaluating metrics for all the scenarios are calculated (shown in Table \ref{table:Expmetrics}), by averaging over all the trajectories.\\
2) All the evaluating metrics are calculated in the measurement region to avoid entry/exit effects.

\subsection{Experimental Observations} \label{sec:benchmark}

\begin{table*}
\centering

\caption{{\footnotesize Evaluating metrics are calculated from the trajectories for all the experimental setups to define a benchmark to evaluate the functioning of pedestrian dynamics models for low pedestrian density scenarios. The experiments show no unsuccessful trajectories, oscillations, or overlaps. Path smoothness, measured by the change in angle between two consecutive motion vectors, ranges from 0 to 2 degrees. The speed deviation from the desired speed at any instant is not more than 23\%. Additionally, the average travel time for volunteers is provided for each case.}}
\label{table:Expmetrics}

\scriptsize
\def\arraystretch{2}
\begin{tabular}{|c|c|c|c|c|c|c|}
\hline
Experimental Setup & Successful Trajectories & Avg. Overlap & Avg. Oscillation & Path Smoothness & Speed Deviation & Travel Time\\\hline

\textbf{SOSP}& 100\% & 0.08\% & 0 & $0.74\pm0.33^\circ$ & $0.05\pm0.03$ & $4.52\pm0.89 s$\\\hline
 
\textbf{Head-On}& 100\% & 0\% & 0 & $0.55\pm0.12^\circ$ & $0.02\pm0.01$ & $3.97\pm0.29 s$\\\hline

 \textbf{Parallel Ped}& 100\% & 0\% & 0 & $0.77\pm0.18^\circ$ & $0.05\pm0.04$ & $5.21\pm1.4 s$\\\hline

 \textbf{MOSP}& & & & & & \\

 \textbf{Case A: 3.74\%}& 100\% & 0\% & 0 & $0.42\pm0.29^\circ$ & $0.05\pm0.03$ & $4.16\pm0.55 s$\\

 \textbf{Case B: 6.54\%}& 100\% & 0\% & 0 & $0.69\pm0.47^\circ$ & $0.11\pm0.08$ & $4.31\pm0.55 s$\\

 \textbf{Case C: 11.22\%}& 100\% & 0.11\% & 1.1E-03 & $1.08\pm0.51^\circ$ & $0.16\pm0.07$ & $4.67\pm0.69 s$\\

 \textbf{Case D: 14.96\%}& 100\% & 0\% & 0 & $0.93\pm0.36^\circ$ & $0.09\pm0.05$ & $4.7\pm0.61 s$\\\hline
 
\end{tabular}

\end{table*}

In this section, we calculate the evaluating metrics defined in Section \ref{sec:evaluatingmetrics} for all experimental setups to establish a benchmark for assessing pedestrian dynamics models in these common scenarios. The values of these metrics are summarized in Table \ref{table:Expmetrics}. Results show no unsuccessful trajectory in any scenario, which shows that the experimental setups, including all cases of MOSP, are such that: a) each pedestrian is able to find a path to their goal without fail, b) pedestrians are not squeezed or overlapped by more than 50\% with other pedestrian or obstacles at any time, c) pedestrians remain within the defined experimental boundaries. Additionally, there are negligible instances of oscillation or overlap observed in any setup during the experiments. Path smoothness was measured by the change in angle between two consecutive motion vectors, and it remained within a low range of 0 to 2 degrees across all setups. This indicates that there were no sudden changes in the direction of motion, even when multiple obstacles were present. The maximum deviation from the desired speed observed among all scenarios was within the range of 24\% of the value, demonstrating that pedestrians tend to maintain a consistent speed regardless of obstructions. Furthermore, the average travel time for volunteers was recorded for each case, providing additional insights into pedestrian dynamics. These evaluation metrics serve as a reliable benchmark for testing the effectiveness of pedestrian dynamics models in these situations.

\section{Simulation results from force-based models}\label{sec:evaluatingForceBasedModels}

In this section, the experimental scenarios are reproduced using five force-based models, maintaining the exact initial and final positions observed during the experiments. Data for the same in each scenario are provided at \url{https://github.com/kanika201293/Pedestrian-Experimental-Data}. The values of the evaluating metrics are calculated from the simulations to assess the models' ability to predict the experimental outcomes. Additionally, a metric for maximum overlap is introduced to provide a more comprehensive evaluation of the models.

In our study, we considered five force-based models to replicate the aforementioned scenarios for the same initial, final positions and desired speeds, as provided by the volunteers. Force-based models represent pedestrians as inertial particles that experience forces on the basis of their surroundings, allowing for the quantification of their behavior. The existing literature identifies three primary forces for motion commonly classified as self-driving force, interaction force, and contact force. The self-driving force enables the pedestrian to reach the goal at some desired speed. A pedestrian ideally should not collide with any other pedestrian or obstacle present in the path. This collision avoidance is fulfilled by a repulsive force, which is termed here as the interaction force. In the case of high pedestrian density, sometimes a pedestrian experiences physical contact with fellow pedestrians or obstacles. The force exerted on the pedestrian due to this physical contact is known as contact force. Contact force is primarily significant in high pedestrian density or panic situations where force experienced by the pedestrian due to physical contact is not under the control of the person. The situations reproduced in this paper are common but at low-to-moderate densities, thereby avoiding the need for contact force. The five force-based models are as follows:\\\\
\textbf{Universal Power Law (UPL): } The UPL is developed by Karamouzas et al. (2014) \cite{karamouzas2014universal}, which is formulated after an empirical analysis of two major crowd datasets. The fundamental concept is that the strength of the interaction force is determined only by a single factor, namely, the time-to-collision, denoted as $\tau$. This term inherently influences the angle-dependent nature of interaction force. Note, interaction with walls utilizes the same formula provided that the pedestrian velocity is directed towards the nearest point to the wall, where the nearest point is treated as a static obstacle with a radius equal to $0\;m$.\\\\
\textbf{Social Force Model - circular (SFMc): } Helbing is one of the earliest researchers to develop a model to study pedestrian dynamics, i.e., Social Force Model \cite{helbing1995social}. SFM has been improved occasionally to bring the simulation results closer to reality. SFMc is a straightforward version of the social force model where the interaction force is only a function of surface-to-surface distance between the pedestrian and the entity. The interaction region of the pedestrian in this case is in the form of a circle \cite{helbing2000simulating}.\\\\
\textbf{Social Force Model - elliptical (SFMe): } One of the earliest versions of social force models used the interaction region of a pedestrian in the form of an ellipse, which is oriented towards the direction of motion \cite{helbing1995social}. In this case, the force is dependent on the distance as well as the velocity of the pedestrian. 

The derived formulae and optimized values of parameters for SFMc and SFMe are taken from the article published by Johansson et al. (2007) \cite{johansson2007specification}.\\\\
\textbf{Centrifugal Force Model - circular (CFMc): }The idea of the centrifugal force model was first proposed by Yu et al. (2005) \cite{yu2005centrifugal} that has a resemblance to the concept of ``centrifugal force". Subsequently, Chraibi expanded upon the model, see \cite{chraibi2009quantitative, chraibi2010generalized}. Similar to SFM, CFM also has a circular and elliptical interaction region distinguished by different definitions of the distance between the pedestrian and the obstacle. For CFMc, the radii of circular potential regions of pedestrians are dependent on the respective velocities. Note that in the case of a stationary pedestrian, the radius of the interaction region takes the minimum value of the actual radius of the entity.\\\\
\textbf{Centrifugal Force Model - elliptical (CFMe): }In this model, the potential region of a pedestrian is elliptical in nature where minor and major axes are dependent on the motion of the pedestrian.\\\\
Note that, for all the models, the simulation time step size $\Delta t$ is set as $0.01\;s$, the maximum time step size that avoids numerical instability (refer to Supplementary material). This study assigns specific colors for each force-based model and the experimental data. The color coding remains consistent throughout the article.

\subsection*{Simulation Results}

\begin{table*}
\centering

\caption{{\footnotesize Comparison of the five force-based models with experimental data using different evaluating metrics for SOSP}}
\label{table:singleObstmetrics}

\scriptsize
\def\arraystretch{2}
\begin{tabular}{|c|c|c|c|c|c|c|c|}
\hline
 & Successful Trajectories & Avg. Oscillation & Avg. Overlap & Max. Overlap & Path Smoothness & Speed Deviation & Travel Time\\\hline
 
 $\textbf{Experiment}$ & \textbf{100\%} & \textbf{0} & \textbf{0.08\%} & \textbf{0.2\%} & $\mathbf{0.74\pm0.33^\circ}$ & $\mathbf{0.05\pm0.03}$ & $\mathbf{4.52\pm0.89 s}$ \\\hline

 $\textbf{UPL}$ & \textbf{100\%} & \textbf{0} & \textbf{0\%} & \textbf{0\%} & $\mathbf{3.33\pm0.88^\circ}$ & $\mathbf{0.06\pm0.04}$ & $\mathbf{4.53\pm0.78 s}$ \\\hline

 $\textbf{SFMc}$ & \textbf{17.86\%} & \textbf{0} &\textbf{35.87\%} & \textbf{100\%} & $\mathbf{0.05\pm0.01^\circ}$ & $\mathbf{0.02\pm0}$ & $\mathbf{4.27\pm0.75 s}$ \\\hline

 $\textbf{SFMe}$ & \textbf{98.21\%} & \textbf{0} & \textbf{2.27\%} & \textbf{52.75\%} & $\mathbf{27.33\pm64.12^\circ}$ & $\mathbf{0.3\pm0.08}$ & $\mathbf{6.27\pm1.68 s}$ \\\hline

$\textbf{CFMc}$ & \textbf{100\%} & \textbf{0} & \textbf{1.29\%} & \textbf{33.1\%} & $\mathbf{0.49\pm0.13^\circ}$ & $\mathbf{0.31\pm0.07}$ & $\mathbf{6.27\pm1.13 s}$ \\\hline

 $\textbf{CFMe}$ & \textbf{100\%} & \textbf{0} & \textbf{0\%} & \textbf{0\%} & $\mathbf{0.45\pm0.2^\circ}$ & $\mathbf{0.34\pm0.08}$ & $\mathbf{6.53\pm1.09 s}$ \\\hline

\end{tabular}
\end{table*}

\textbf{SOSP: }As mentioned earlier, in this case, a pedestrian aims to reach a goal by avoiding a stationary obstacle placed in the middle of the path. Figure \ref{fig:experimentalSetups}(a) illustrates the initial position of pedestrian A (taken as (0, 0)). In real-life situations, it is unlikely that a person stands exactly at the expected position without any offset. During the experiment, we observed that 58 participants had an average offset of 0.09 m with a standard deviation of 0.07 m from their indicated initial position (0, 0). The participants' average desired speed (speed without any obstacle) was 1.47 m/s with a standard deviation of 0.16 m/s.

\begin{figure}
    \centering
    \subfigure[]
    {
        \includegraphics[width= 1\linewidth]{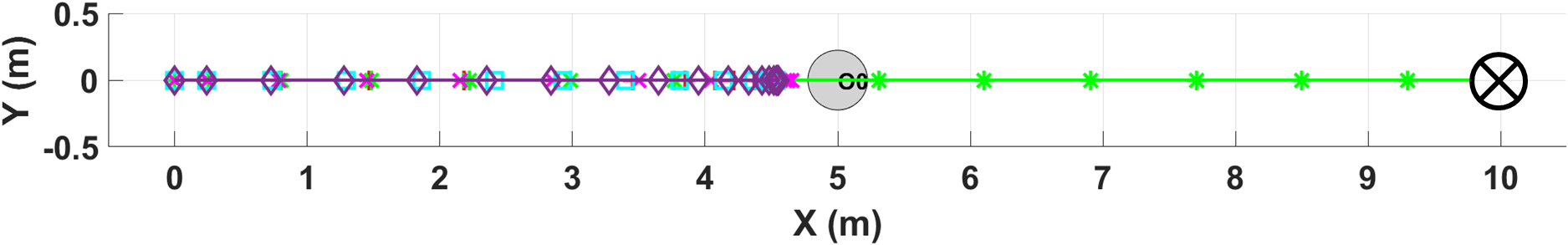}
        \label{fig:singleObst_noOffset_traj}
    }
    
    \subfigure[]
    {
        \includegraphics[width= 1\linewidth]{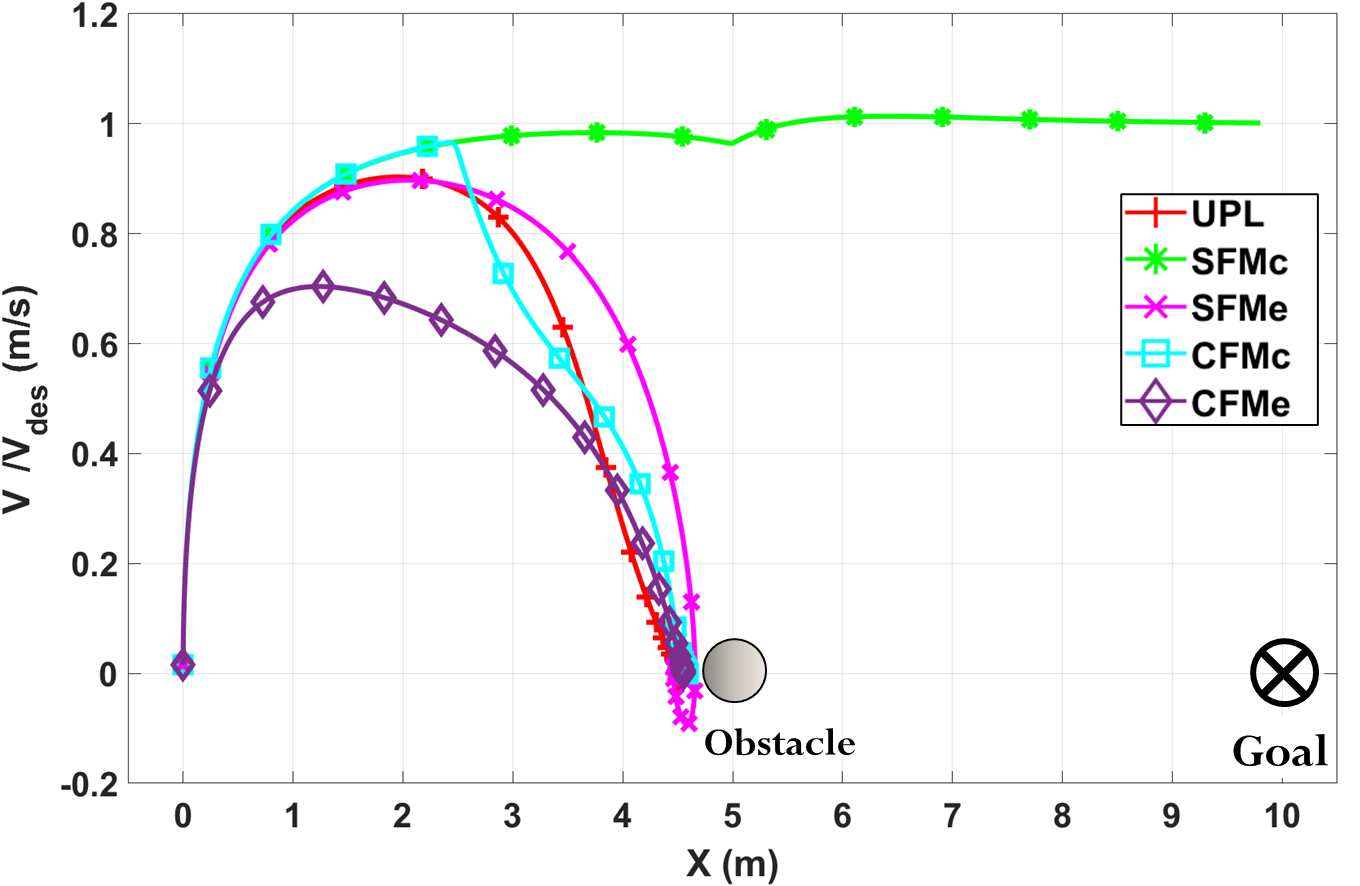}
        \label{fig:singleObst_noOffset_vel}
    }

    \caption{{\footnotesize  (a) Trajectories for all models show overlap, showing no transverse displacement due to collinear forces. Pedestrians get stuck before the obstacle in all models except for SFMc, where pedestrian can pass through the obstacle due to low repulsion. (b) Normalized speed plotted against the x-component of the pedestrian's position. Speeds for all models approach 0 m/s, indicating pedestrians get stuck, except for SFMc. SFMe exhibits oscillation before getting stuck, indicated by negative velocity.}} \label{fig:singleObst_noOffset}
    
\end{figure}

\begin{figure}
    \centering
    \subfigure[Trajectory graphs (Multimedia available online)]
    {
        \includegraphics[width= 0.8\linewidth]{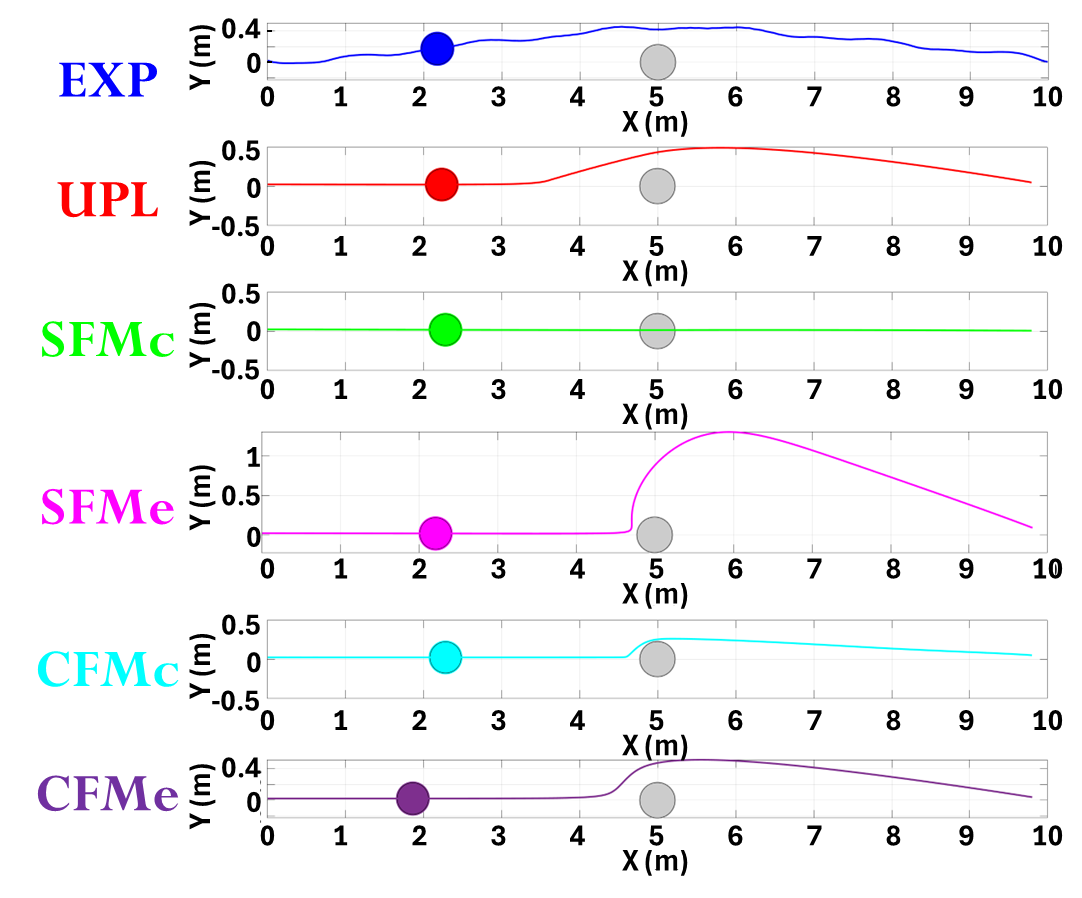}
        \label{fig:singleObst_regen_traj}
    }
    
    \subfigure[Normalized speed graph]
    {
        \includegraphics[width= 0.8\linewidth]{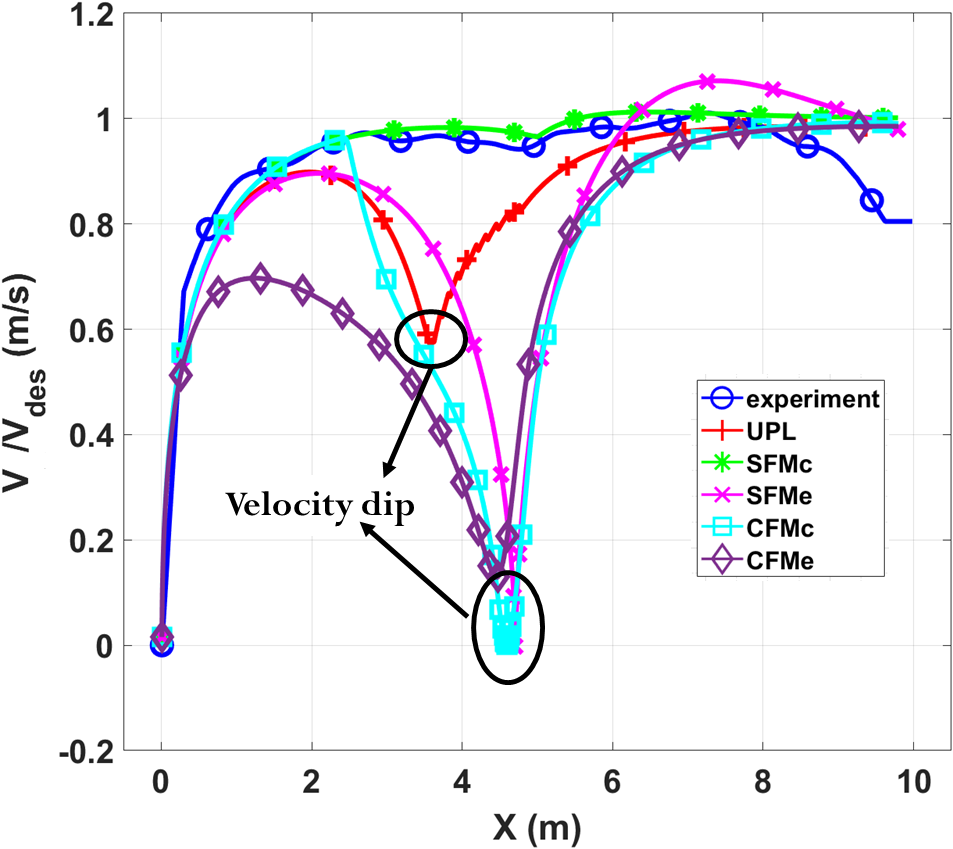}
        \label{fig:singleObst_regen_vel}
    }

    \caption{{\footnotesize  SOSP: (a) Trajectories and (b) Normalized speed for experimental and simulated trajectories. The experimental data shows anticipatory direction changes with minimal speed change to avoid the obstacle. In contrast, simulated models, except SFMc, exhibit sudden direction changes near the obstacle, resulting in velocity dips. Overlaps with the obstacle are noted for SFMc, SFMe, and CFMc, with no oscillation or backward motion ($V < 0m/s$) observed in any model.}} \label{fig:singleObst_regen}
    
\end{figure}

Assuming a pedestrian starts from the exact initial position so that the pedestrian, the obstacle, and the goal are collinear, as shown in the schematic diagram, simulations using the UPL model predicts the pedestrian getting stuck after reaching near the obstacle positioned in the middle (refer to Figure \ref{fig:singleObst_noOffset_traj}). The collinearity of the pedestrian, obstacle, and goal results in attractive and repulsive forces, also collinear but opposite in direction. As the pedestrian approaches the obstacle, the repulsive force increases, counterbalancing the self-driving force. Due to the absence of any transverse force, the pedestrian cannot change its direction and remains stuck. The same problem of collinearity is also observed in the case of all other force-based models considered in this study.

Figure \ref{fig:singleObst_noOffset_traj} shows that the trajectories for all models overlap, indicating no transverse displacement due to collinear forces. The pedestrian gets stuck just before the obstacle in all models except SFMc, where overlap with the obstacle is observed, allowing the pedestrian to reach the goal without deviation. In SFMc, the collinearity problem results in no transverse movement, and the low repulsive force of the model leads to overlap. The normalized speed is shown against the x-component of the pedestrian's position in Figure \ref{fig:singleObst_noOffset_vel}. We observe that the speeds for all models approach 0, indicating that pedestrians get stuck, except for SFMc. As discussed earlier, due to the low repulsive force magnitude, the pedestrian continues with overlap with no decrement in speed.

However, in real-life scenarios, it is unlikely that a person stands precisely in the expected position with no offset. Trajectories and normalized speed profiles for one of the data are plotted in Figure \ref{fig:singleObst_regen} (multimedia available online). In this case, the volunteer in the experiment started with an offset as low as 11 mm, yet there were no sudden changes in the trajectory or speed during the experiment, unlike simulation results. As explained next, all models showed significant slowdown of the pedestrian as it approached the obstacle in the simulations, which is in sharp contrast to the experimental observations. The possible reason behind such behavior is force-based models relying on forces to navigate around the obstacles, which naturally leads to a change in the magnitude of the velocity (speed), as can be seen in Figure \ref{fig:singleObst_regen_vel}.

While simulating the experimental scenarios using force-based models using the same initial and final positions, we calculated and reported the corresponding evaluating metrics in Table \ref{table:singleObstmetrics}. Model SFMc predicted over 80\% unsuccessful trajectories, SFMe showed less than 2\% unsuccessful trajectories, while all other models had all successful trajectories. The large number of unsuccessful trajectories for model SFMc is primarily due to high overlap, with a maximum overlap of 100\%. This high failure rate stems from SFMc's weak interaction force, leading to an inability to avoid obstacles effectively, as depicted in Figure \ref{fig:singleObst_regen_traj} (multimedia available online). Some overlap was observed in models SFMe and CFMc, but no overlap occurred with UPL and CFMe. No backward motion or oscillation was reported for any model. However, sudden changes in direction were noted under the `path smoothness' metric, with model SFMe showing a maximum change exceeding 90 degrees, as illustrated in the trajectory. It is important to note that human walking is inherently less smooth than simulated motion, which can result in lower path smoothness values in simulations compared to experimental values in some cases. Large speed deviations from the desired speed were observed in models SFMe, CFMc, and CFMe, evident in the velocity dips highlighted in Figure \ref{fig:singleObst_regen_vel}. These speed slowdowns significantly increased travel time for these three models, and respective values are provided in Table \ref{table:singleObstmetrics}.
\begin{table*}
\centering

\caption{{\footnotesize Comparison of the five force-based models with experimental data using different evaluating metrics for Head-On.}}
\label{table:headOnmetrics}

\scriptsize
\def\arraystretch{2}
\begin{tabular}{|c|c|c|c|c|c|c|c|}
\hline
 & Successful Trajectories & Avg. Oscillation & Avg. Overlap & Max. Overlap & Path Smoothness & Speed Deviation & Travel Time\\\hline
 
$\textbf{Experiment}$ & \textbf{100\%} & \textbf{0} & \textbf{0\%} & \textbf{0\%} & $\mathbf{0.55\pm0.12^\circ}$ & $\mathbf{0.02\pm0.01}$ & $\mathbf{3.97\pm0.29 s}$ \\\hline

$\textbf{UPL}$ & \textbf{100\%} & \textbf{0} & \textbf{0\%} & \textbf{0\%} & $\mathbf{2.81\pm0.44^\circ}$ & $\mathbf{0.03\pm0.03}$ & $\mathbf{3.99\pm0.31 s}$ \\\hline

$\textbf{SFMc}$ & \textbf{52.38\%} & \textbf{0} &\textbf{24.68\%} & \textbf{97.8\%} & $\mathbf{0.05\pm0.01^\circ}$ & $\mathbf{0.01\pm0.0}$ & $\mathbf{3.89\pm0.26 s}$ \\\hline

$\textbf{SFMe}$ & \textbf{76.19\%} & \textbf{0} & \textbf{11.4\%} & \textbf{81.66\%} & $\mathbf{10.03\pm38.29^\circ}$ & $\mathbf{0.11\pm0.03}$ & $\mathbf{4.41\pm0.45 s}$ \\\hline

$\textbf{CFMc}$ & \textbf{100\%} & \textbf{0} & \textbf{0.02\%} & \textbf{0.64\%} & $\mathbf{0.59\pm0.23^\circ}$ & $\mathbf{0.28\pm0.07}$ & $\mathbf{5.45\pm0.69 s}$ \\\hline

$\textbf{CFMe}$ & \textbf{80.95\%} & \textbf{0} & \textbf{9.55\%} & \textbf{74.2\%} & $\mathbf{8.62\pm4.73^\circ}$ & $\mathbf{0.38\pm0.06}$ & $\mathbf{6.08\pm0.44 s}$ \\\hline

\end{tabular}
\end{table*}

\begin{figure} [h]
    \centering
    \subfigure[Trajectory graphs (Multimedia available online)]
    {
        \includegraphics[width= 0.8\linewidth]{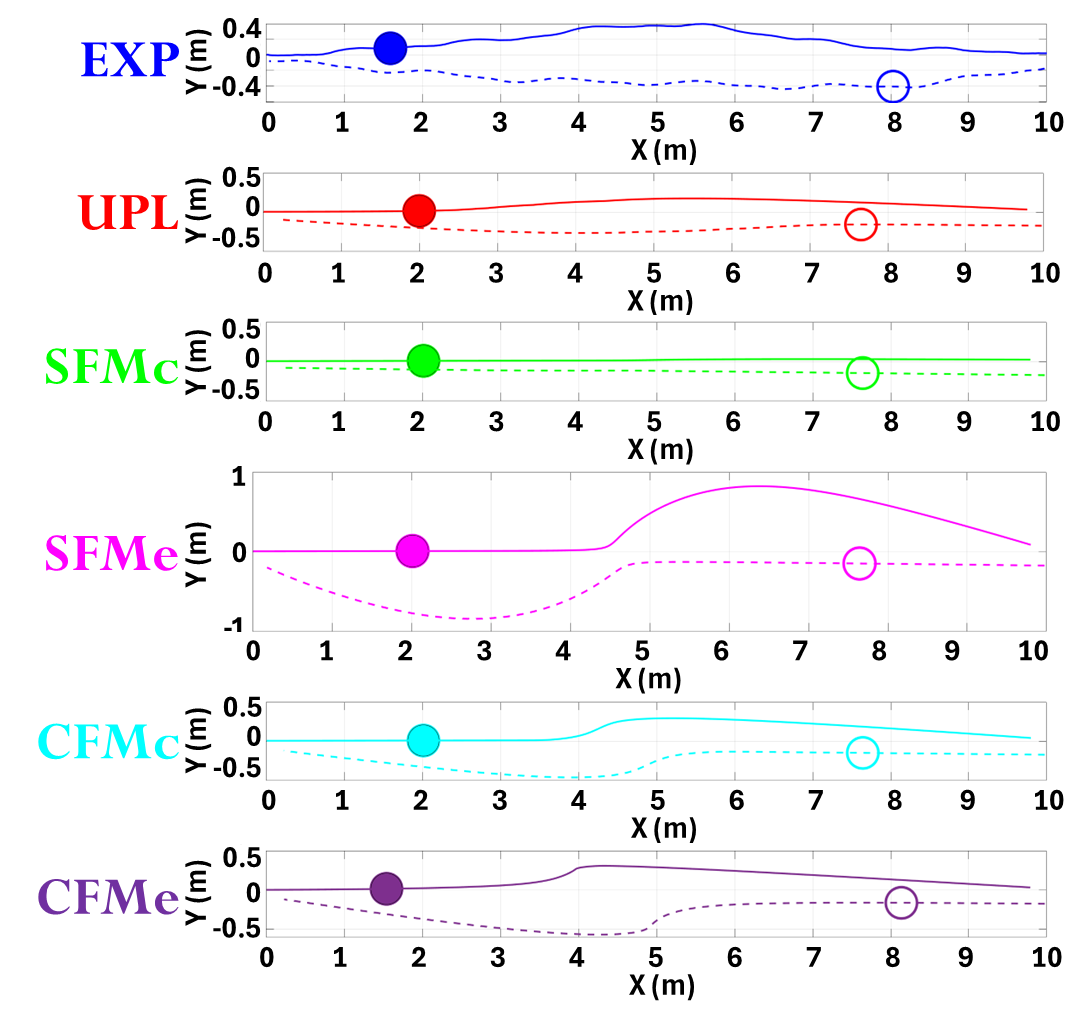}
        \label{fig:headOn_traj}
    }
    
    \subfigure[Normalized speed (Ped A) graph]
    {
        \includegraphics[width= 0.8\linewidth]{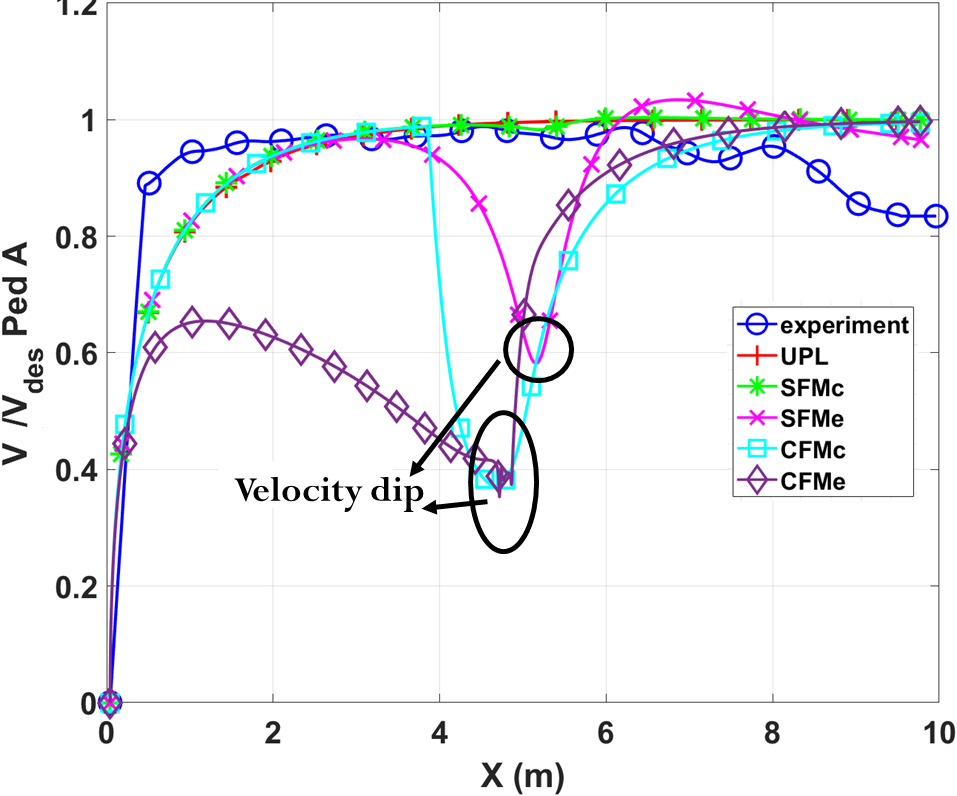}
        \label{fig:headOn_vxVSx}
    }

    \caption{{\footnotesize  Head-On: (a) Trajectories and (b) Normalized speed (Ped A) graphs are presented for both experimental and simulated data. The experimental data indicates an anticipatory change in direction to avoid collisions, with minimal speed fluctuations. In the trajectory plots, solid lines with filled circles represent pedestrian A, while dashed lines with empty circles represent pedestrian B. The velocity profiles show a dip when avoiding collisions in all models except SFMc.}} \label{fig:headOn}
    
\end{figure}

An evaluation of force-based models for SOSP revealed that model SFMc had less than 20\% successful trajectories due to weak interaction forces resulting in high overlap. In contrast, models SFMe had only a few unsuccessful trajectories, while UPL, CFMc, and CFMe had none. Additionally, models UPL and SFMe exhibited sudden changes in path direction, and models SFMe, CFMc, and CFMe experienced significant speed deviations, which led to increased travel times. Note, nearly all models failed when the pedestrian, obstacle, and goal are collinear with zero offset, as discussed earlier.

\textbf{Head-On: } One example from the head-on experiments is plotted in Figure \ref{fig:headOn} (multimedia available online). In trajectory plots, solid lines with filled circles represent pedestrian A, and dashed lines with empty circles denote pedestrian B. Table \ref{table:headOnmetrics} reports the evaluating metrics for head-on scenarios simulated using force-based models. Unsuccessful cases are predicted for models SFMc, SFMe, and CFMe, with high overlapping proportions. For SFMc and SFMe, the overlap is also evident from the trajectories shown in Figure \ref{fig:headOn_traj} (multimedia available online). Models UPL and CFMc report 0\% and less than 1\% maximum overlap, respectively, and no oscillation is observed in any model. Large sudden deviations in the direction of motion are observed in models SFMe and CFMe, also shown in the trajectory. The metric of path smoothness measures the sudden change in the direction of motion rather than the overall change in the path. Hence, the value of path smoothness is lower for CFMc relative to CFMe. Although the overall deviation is large for CFMc, it is smoother compared to the abrupt changes observed in CFMe's trajectory at around $x = 4\;m$ in Figure \ref{fig:headOn_traj} (multimedia available online). Significant speed deviations are reported for models CFMc and CFMe, as can also be seen in the velocity profiles of Figure \ref{fig:headOn_vxVSx}. Consequently, these models also exhibit longer travel times, as indicated in Table \ref{table:headOnmetrics}. For other models, the simulation predicts travel times similar to the experiments.

\begin{table*}
\centering

\caption{{\footnotesize Comparison of the five force-based models with experimental data using different evaluating metrics for Parallel-Ped.}}
\label{table:parallelPedmetrics}

\scriptsize
\def\arraystretch{2}
\begin{tabular}{|c|c|c|c|c|c|c|c|}
\hline
 & Successful Trajectories & Avg. Oscillation & Avg. Overlap & Max. Overlap & Path Smoothness & Speed Deviation & Travel Time\\\hline
 
$\textbf{Experiment}$ & \textbf{100\%} & \textbf{0} & \textbf{0\%} & \textbf{0\%} & $\mathbf{0.77\pm0.18^\circ}$ & $\mathbf{0.05\pm0.04}$ & $\mathbf{5.21\pm1.4 s}$ \\\hline

$\textbf{UPL}$ & \textbf{100\%} & \textbf{0} & \textbf{0\%} & \textbf{0\%} & $\mathbf{3.03\pm1.08^\circ}$ & $\mathbf{0.02\pm0.01}$ & $\mathbf{5.0\pm1.2 s}$ \\\hline

$\textbf{SFMc}$ & \textbf{66.67\%} & \textbf{0} &\textbf{15.61\%} & \textbf{95.25\%} & $\mathbf{0.04\pm0.02^\circ}$ & $\mathbf{0.04\pm0.01}$ & $\mathbf{5.01\pm1.2 s}$ \\\hline

$\textbf{SFMe}$ & \textbf{100\%} & \textbf{0} & \textbf{0\%} & \textbf{0\%} & $\mathbf{0.03\pm0.03^\circ}$ & $\mathbf{0.22\pm0.05}$ & $\mathbf{4.71\pm0.26 s}$ \\\hline

$\textbf{CFMc}$ & \textbf{100\%} & \textbf{0} & \textbf{0\%} & \textbf{0\%} & $\mathbf{0.05\pm0.05^\circ}$ & $\mathbf{0.19\pm0.1}$ & $\mathbf{5.56\pm0.83 s}$ \\\hline

$\textbf{CFMe}$ & \textbf{100\%} & \textbf{0} & \textbf{0\%} & \textbf{0\%} & $\mathbf{0.04\pm0.05^\circ}$ & $\mathbf{0.18\pm0.15}$ & $\mathbf{5.93\pm0.57}$ \\\hline

\end{tabular}
\end{table*}

\begin{figure*}
    \includegraphics[width= 0.8\linewidth]{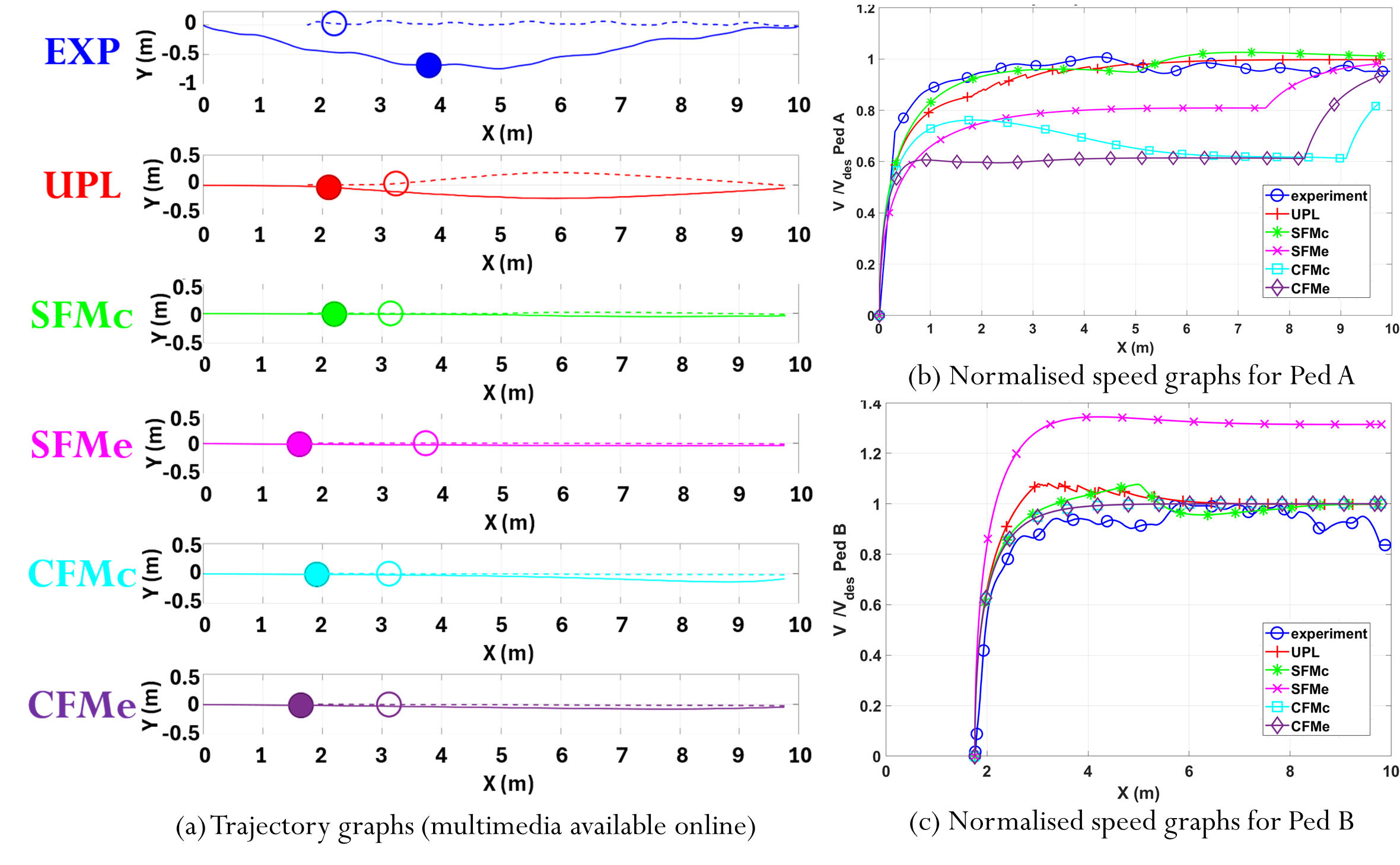}
    \caption{{\footnotesize 
    ParallelPed: (a) Trajectories (Multimedia available online) and (b) normalized speed graphs are presented for both experimental and simulated data. In the experiment, pedestrian A (faster, solid line with filled circles) avoids pedestrian B (slower, dashed line with empty circles) by changing direction, while B continues without interruption. Both show minimal speed changes. In simulations, A is unable to overtake B, resulting in straight-line trajectories in all models except UPL. Velocity profiles indicate A fails to reach its desired speed in SFMe, CFMc, and CFMe models, while B maintains consistent speed, with B experiencing a push effect in SFMe.}} 
    \label{fig:parallelPed_reg}
\end{figure*}

An evaluation of force-based models for the head-on simulations reveal that models SFMc and SFMe predict less than 80\% successful trajectories, characterized by high overlap. Model CFMe has a few unsuccessful trajectories, while models UPL and CFMc had none. Furthermore, sudden changes in path direction were observed in models UPL, CFMe, and SFMe, and significant speed deviations were noted in models SFMe, CFMc, and CFMe, contributing to increased travel times. Note, similar to the previous case, models also fail for collinear pedestrians in Head-On simulations, in contrast to experiments.

\begin{table*}
\centering

\caption{{\footnotesize Comparison of the five force-based models with experimental data using different evaluating metrics for MOSP.}}
\label{table:MOSPmetrics}

\scriptsize
\def\arraystretch{2}
\begin{tabular}{|c|c|c|c|c|c|c|c|}
\hline
$\textbf{MOSP}$ & Successful Trajectories & Avg. Oscillation & Avg. Overlap & Max. Overlap & Path Smoothness & Speed Deviation & Travel Time\\\hline

\multicolumn{1}{c}{\textbf{Case A: 03.74\%}} & \multicolumn{7}{c}{}\\\hline
 
$\textbf{Experiment}$ & \textbf{100\%} & \textbf{0} & \textbf{0\%} & \textbf{0\%} & $\mathbf{0.42\pm0.29^\circ}$ & $\mathbf{0.05\pm0.03}$ & $\mathbf{4.16\pm0.55 s}$ \\\hline

$\textbf{UPL}$ & \textbf{100\%} & \textbf{0.01} & \textbf{0\%} & \textbf{0\%} & $\mathbf{2.58\pm1.5^\circ}$ & $\mathbf{0.07\pm0.06}$ & $\mathbf{4.23\pm0.51 s}$ \\\hline

$\textbf{SFMc}$ & \textbf{61.51\%} & \textbf{0} &\textbf{18.3\%} & \textbf{100\%} & $\mathbf{0.04\pm0.01^\circ}$ & $\mathbf{0.01\pm0.0}$ & $\mathbf{3.96\pm0.45 s}$ \\\hline

$\textbf{SFMe}$ & \textbf{83.26\%} & \textbf{0.07} & \textbf{0.42\%} & \textbf{27.4\%} & $\mathbf{29.04\pm64.65^\circ}$ & $\mathbf{0.42\pm0.18}$ & $\mathbf{7.34\pm3.44 s}$ \\\hline

$\textbf{CFMc}$ & \textbf{100\%} & \textbf{0} & \textbf{1.6\%} & \textbf{39.5\%} & $\mathbf{0.56\pm0.51^\circ}$ & $\mathbf{0.38\pm0.14}$ & $\mathbf{6.73\pm1.7 s}$ \\\hline

$\textbf{CFMe}$ & \textbf{92.05\%} & \textbf{0} & \textbf{6.19\%} & \textbf{100\%} & $\mathbf{1.27\pm10.83^\circ}$ & $\mathbf{0.41\pm0.12}$ & $\mathbf{7.07\pm2.03}$ \\\hline

\hline
\multicolumn{1}{c}{\textbf{Case B: 06.54\%}} & \multicolumn{7}{c}{}\\\hline
 
$\textbf{Experiment}$ & \textbf{100\%} & \textbf{0} & \textbf{0\%} & \textbf{0\%} & $\mathbf{0.69\pm0.47^\circ}$ & $\mathbf{0.11\pm0.08}$ & $\mathbf{4.31\pm0.55 s}$ \\\hline

$\textbf{UPL}$ & \textbf{100\%} & \textbf{0.02} & \textbf{0\%} & \textbf{0\%} & $\mathbf{5.46\pm5.07^\circ}$ & $\mathbf{0.18\pm0.15}$ & $\mathbf{4.9\pm1.41 s}$ \\\hline

$\textbf{SFMc}$ & \textbf{26.06\%} & \textbf{0} &\textbf{27.35\%} & \textbf{100\%} & $\mathbf{0.05\pm0.01^\circ}$ & $\mathbf{0.03\pm0.0}$ & $\mathbf{3.86\pm0.52 s}$ \\\hline

$\textbf{SFMe}$ & \textbf{31.38\%} & \textbf{0.07} & \textbf{7.31\%} & \textbf{68.2\%} & $\mathbf{108.31\pm86.9^\circ}$ & $\mathbf{0.75\pm0.24}$ & $\mathbf{13.74\pm6.99 s}$ \\\hline

$\textbf{CFMc}$ & \textbf{87.77\%} & \textbf{0} & \textbf{2.29\%} & \textbf{50.73\%} & $\mathbf{12.88\pm37.51^\circ}$ & $\mathbf{0.57\pm0.14}$ & $\mathbf{9.71\pm4.08 s}$ \\\hline

$\textbf{CFMe}$ & \textbf{96.81\%} & \textbf{0} & \textbf{1.27\%} & \textbf{97.23\%} & $\mathbf{3.3\pm17.37^\circ}$ & $\mathbf{0.58\pm0.07}$ & $\mathbf{7.07\pm2.03}$ \\\hline

\hline
\multicolumn{1}{c}{\textbf{Case C: 11.22\%}} & \multicolumn{7}{c}{}\\\hline
 
$\textbf{Experiment}$ & \textbf{100\%} & \textbf{1.1E-03} & \textbf{0.11\%} & \textbf{0.18\%} & $\mathbf{1.08\pm0.51^\circ}$ & $\mathbf{0.16\pm0.07}$ & $\mathbf{4.67\pm0.69 s}$ \\\hline

$\textbf{UPL}$ & \textbf{60.33\%} & \textbf{0.01} & \textbf{0\%} & \textbf{8.52\%} & $\mathbf{82.54\pm77.23^\circ}$ & $\mathbf{0.58\pm0.3}$ & $\mathbf{12.15\pm6.8 s}$ \\\hline

$\textbf{SFMc}$ & \textbf{3.8\%} & \textbf{0} &\textbf{31.53\%} & \textbf{100\%} & $\mathbf{0.06\pm0.01^\circ}$ & $\mathbf{0.03\pm0.0}$ & $\mathbf{3.92\pm0.52 s}$ \\\hline

$\textbf{SFMe}$ & \textbf{0\%} & \textbf{0.01} & \textbf{1\%} & \textbf{77.45\%} & $\mathbf{NA}$ & $\mathbf{NA}$ & $\mathbf{NA}$ \\\hline

$\textbf{CFMc}$ & \textbf{84.24\%} & \textbf{0} & \textbf{6.23\%} & \textbf{100\%} & $\mathbf{7.56\pm27.4^\circ}$ & $\mathbf{0.7\pm0.08}$ & $\mathbf{13.46\pm2.86 s}$ \\\hline

$\textbf{CFMe}$ & \textbf{11.95\%} & \textbf{0} & \textbf{26.44\%} & \textbf{100\%} & $\mathbf{73.58\pm65.29^\circ}$ & $\mathbf{0.74\pm0.15}$ & $\mathbf{14.89\pm3.79}$ \\\hline

\hline
\multicolumn{1}{c}{\textbf{Case D: 14.96\%}} & \multicolumn{7}{c}{}\\\hline
 
$\textbf{Experiment}$ & \textbf{100\%} & \textbf{0} & \textbf{0\%} & \textbf{0\%} & $\mathbf{0.93\pm0.36^\circ}$ & $\mathbf{0.09\pm0.05}$ & $\mathbf{4.7\pm0.61 s}$ \\\hline

$\textbf{UPL}$ & \textbf{56.52\%} & \textbf{0.01} & \textbf{0\%} & \textbf{0.99\%} & $\mathbf{90.3\pm79.66^\circ}$ & $\mathbf{0.62\pm0.29}$ & $\mathbf{13.74\pm7.35 s}$ \\\hline

$\textbf{SFMc}$ & \textbf{21.01\%} & \textbf{0} &\textbf{28.92\%} & \textbf{100\%} & $\mathbf{0.07\pm0.01^\circ}$ & $\mathbf{0.04\pm0.0}$ & $\mathbf{4.3\pm0.44 s}$ \\\hline

$\textbf{SFMe}$ & \textbf{0\%} & \textbf{0} & \textbf{0\%} & \textbf{0\%} & $\mathbf{NA}$ & $\mathbf{NA}$ & $\mathbf{NA}$ \\\hline

$\textbf{CFMc}$ & \textbf{72.46\%} & \textbf{0} & \textbf{13.57\%} & \textbf{100\%} & $\mathbf{13.06\pm37.16^\circ}$ & $\mathbf{0.69\pm0.15}$ & $\mathbf{14.94\pm4.13 s}$ \\\hline

$\textbf{CFMe}$ & \textbf{6.88\%} & \textbf{0} & \textbf{33.76\%} & \textbf{100\%} & $\mathbf{47.83\pm62.57^\circ}$ & $\mathbf{0.71\pm0.17}$ & $\mathbf{14.18\pm4.31}$ \\\hline

\end{tabular}

\end{table*} 

\textbf{Parallel-Ped: }In this scenario, two pedestrians, A and B, share a common goal, as illustrated in Figure \ref{fig:experimentalSetups}(c). One example of speed profiles and trajectories for both pedestrians generated using force-based models are provided in Figure \ref{fig:parallelPed_reg} (multimedia available online). In trajectory plots, solid lines with filled circles represent pedestrian A, while dashed lines with empty circles represent pedestrian B. Evaluating metrics obtained from simulations are shown in Table \ref{table:parallelPedmetrics}. The model SFMc predicted more than 30\% unsuccessful trajectories due to large overlap in most cases, with a maximum overlap of 95.25\%. For scenarios SOSP, head-on, and parallel ped, involving only two entities, the main cause of unsuccessful trajectories is overlap rather than the pedestrian getting stuck. Oscillations are not predicted in any model for this scenario. The trajectories of pedestrian A in Figure \ref{fig:parallelPed_reg}(a) (multimedia available online) indicate that, for most models, the faster pedestrian A is unable to overtake the slower pedestrian B, resulting in mostly straight-line trajectories. However, in experiments, pedestrian A overtakes B, in contrast to these simulated predictions. Thus, path smoothness for most models is better than the experimental value, except for UPL. In the UPL model, pedestrian A successfully overtakes pedestrian B. However, pedestrian B also changes the path direction to allow passage for pedestrian A, unlike during the experiment where A finds its own path without affecting B's motion. This is logical as pedestrian B, positioned ahead of A, cannot see A and should not alter its path. For UPL, sudden direction change for both pedestrians ranges from 2 to 4 degrees. Models SFMe, CFMc, and CFMe exhibit significant speed deviations from the desired speed, with maximum deviations between 20\% and 48\%. Large deviation in velocities from the desired speed ($V_{des}$) is also observed in the normalized speed plots given in Figure \ref{fig:parallelPed_reg}(b)-(c). For these models, pedestrian A cannot overtake B, resulting in a slower speed of A (Figure \ref{fig:parallelPed_reg}(b)), while the speed of B is maintained for models CFMc and CFMe (Figure \ref{fig:parallelPed_reg}(c)). The SFMe model shows an unrealistic push effect on pedestrian B, increasing the speed by up to 1.4 times. The models predicted travel times similar to the experiment.

An evaluation of force-based models for parallel pedestrian scenarios revealed that SFMc had fewer than 80\% successful trajectories and exhibited significant overlap, whereas all other models had 100\% successful trajectories. Sudden changes in path direction were noted exclusively in model UPL, while significant speed deviations were observed in models SFMe, CFMc, and CFMe, leading to increased travel times. Apart from UPL, all models predicted A being behind B and getting slowed down, in contrast to experiments. However, as noted earlier, UPL predicts a path change of B, which contradicts the experiments.

\textbf{MOSP: }In the MOSP scenario, a single pedestrian passes through a maze of randomly placed obstacles to reach the goal, with experiments conducted across four area fractions: 3.74\% (Case A), 6.54\% (Case B), 11.22\% (Case C), and 14.96\% (Case D). Evaluating metrics calculated by reproducing each setup for the exact initial and final conditions are reported in Table \ref{table:MOSPmetrics}. These show that as the area fraction covered by the obstacles increases, the number of unsuccessful trajectories also increases. While it is commonly reported in the literature that force-based models perform well in complex scenarios such as lane formation under moderate to high densities, our findings indicate a decline in model performance as density increases. However, one difference in our situations considered here is the aspect of navigation around a crowded zone with static obstacles, unlike the dynamics in a crowd where every pedestrian is mobile.

Note, such situations of pedestrians walking around several other stationary obstacles (or other standing pedestrians) are extremely common across crowded public spaces, like railway stations, bus stations, food joints etc. In MOSP, the large number of obstacles leads to unsuccessful trajectories due to a combination of three factors mentioned in Section \ref{sec:evaluatingmetrics}. For the SFMe model, no successful trajectories were observed in Cases C and D, as high repulsive forces prevented pedestrians from entering the measurement region, resulting in `Not Attempted (NA)' for these cases, (refer to Appendix Figure B-5(b)). Surprisingly, the SFMc model showed an increase of successful trajectories from Case C to D due to the obstacle configuration favorable for the model, allowing the pedestrian to reach the goal with low overlap, even when there is a minimal deviation in path due to the weak repulsive forces of the model (refer to Supplementary material Figure B-5(a)). Both average and maximum overlaps increased with area fraction. However, this metric is influenced by obstacle configuration, pedestrian position, and the number of interacting obstacles. Thus, this trend should not be seen as a rule. Insignificant oscillation or backward motion was observed across all models. As the number of obstacles increased, the number of turns required to pass through the maze also increased, leading to higher sudden changes in motion direction, reflected in large path smoothness values. Due to frequent changes in path direction, speed deviation from the desired speed also increases with the area fraction. Normalized speed graphs for the four cases, averaged over all the trajectories, can be observed in Supplementary material Figure B-4. However, the SFMc model reported low path smoothness and speed deviation, resulting in increased travel time with area fraction for all models except SFMc. However, note that the proportion of successful trajectories with SFMc is pretty low ($<25\%$) for both area fractions.

In the MOSP scenario, the study reveals that as obstacle density increases, the frequency of unsuccessful trajectories rises, with no model achieving an 80\% success rate at the density of 14.96\%. The pedestrian models exhibit widely divergent responses to these challenges: the SFMe model completely fails in higher-density conditions, whereas the SFMc model demonstrates improved success rates, along with reduced path smoothness and speed deviation, due to favorable obstacle configurations for the model. The analysis also shows that with an increase in obstacles, there is a corresponding rise in path complexity, speed deviation, and travel time. However, these trends are not universally applicable, as they depend on the specific configurations of obstacles and the positioning of pedestrians. As discussed, this is a common scenario in any public gathering/space, where one person may need to navigate around obstacles and other people to reach a different location. For such an everyday scenario, simulation models predicting unsuccessful trajectories reveal serious shortcomings.

\section{Evaluation of Models}

The evaluation procedure involves two stages:

\begin{itemize}
    \item Stage \Romannum{1}: The models must meet an eligibility criterion of at least 80\% successful trajectories (Table \ref{table:stage1}).
    \item Stage \Romannum{2}: The shortlisted models are then evaluated using a scoring system, where a model's score is determined by the number of times its evaluating metrics fall within the specified cutoff values (Table \ref{table:stage2}, \ref{table:scoring}).
\end{itemize}

\begin{table}
\centering

\caption{{\footnotesize Stage \Romannum{1}: Models meeting eligibility criteria are reported in this table. For each experimental setup, models achieving more than 80\% successful cases are marked as $\checkmark$, otherwise ×. Models with the number of $\checkmark$ greater than or equal to the number of × are shortlisted for further evaluation in Stage \Romannum{2}, given in Table \ref{table:stage2}.}}
\label{table:stage1}

\scriptsize
\def\arraystretch{2}
\begin{tabular}{|c|c|c|c|c|c|}
\hline
 & $\textbf{UPL}$ & $\textbf{SFMc}$ & $\textbf{SFMe}$ & $\textbf{CFMc}$ & $\textbf{CFMe}$\\\hline
 
 $\textbf{SOSP}$ & $\checkmark$ & $\times$ & $\checkmark$ & $\checkmark$ & $\checkmark$\\\hline

 $\textbf{Head-On}$ & $\checkmark$ & $\times$ & $\times$ & $\checkmark$ & $\checkmark$\\\hline

 $\textbf{Parallel Ped}$ & $\checkmark$ & $\times$ & $\checkmark$ & $\checkmark$ & $\checkmark$\\\hline

 $\textbf{MOSP:}$ & & & & & \\

 $\textbf{Case A(03.74\%)}$ & $\checkmark$ & $\times$ & $\checkmark$ & $\checkmark$ & $\checkmark$\\

 $\textbf{Case B(06.54\%)}$ & $\checkmark$ & $\times$ & $\times$ & $\checkmark$ & $\checkmark$\\

 $\textbf{Case C(11.22\%)}$ & $\times$ & $\times$ & $\times$ & $\checkmark$ & $\times$\\

 $\textbf{Case D(14.96\%)}$ & $\times$ & $\times$ & $\times$ & $\times$ & $\times$\\\hline

 & $\textbf{Elig.}$ & $\textbf{NE}$ & $\textbf{NE}$ & $\textbf{Elig.}$ & $\textbf{Elig.}$\\\hline

\end{tabular}
\end{table}

\begin{table*}
\centering

\caption{{\footnotesize Stage \Romannum{2}: Labels $\checkmark$ and × are assigned to the shortlisted models from Stage \Romannum{1}, based on whether their average evaluating metric values fall within the specified cutoff values or not, respectively. The cutoff values are mentioned in the first column corresponding to each metric. The values for overlap and oscillation metrics are set as 10\% and 0.005, respectively, for each experimental setup. The cutoff values are determined using the $3 \sigma$ rule for the other three metrics, adding three standard deviations to the mean values observed during the experiments. The final scoring for the models is provided in Table \ref{table:scoring}}}
\label{table:stage2}

\scriptsize
\def\arraystretch{2}
\begin{tabular}{|c|c|c|c|c|c|c|c|c|c|c|c|c|c|c|c|c|c|c|c|c|}
\hline
\multicolumn{1}{c}{\begin{scriptsize}Metrics $\boldsymbol{\rightarrow}$\end{scriptsize}} & \multicolumn{4}{c}{\begin{small}\textbf{Avg. Overlap}\end{small}} & \multicolumn{4}{c}{\begin{small}\textbf{Avg. Oscillation}\end{small}} & \multicolumn{4}{c}{\begin{small}\textbf{Path Smoothness}\end{small}} & \multicolumn{4}{c}{\begin{small}\textbf{Speed Deviation}\end{small}} & \multicolumn{4}{c}{\begin{small}\textbf{Travel Time}\end{small}}\\\hline

\begin{scriptsize}Exp. Setup $\boldsymbol{\downarrow}$\end{scriptsize} & \begin{tiny}Cutoff\end{tiny} & \begin{tiny}UPL\end{tiny} & \begin{tiny}CFMc\end{tiny} & \begin{tiny}CFMe\end{tiny} & \begin{tiny}Cutoff\end{tiny} & \begin{tiny}UPL\end{tiny} & \begin{tiny}CFMc\end{tiny} & \begin{tiny}CFMe\end{tiny} & \begin{tiny}Cutoff\end{tiny} & \begin{tiny}UPL\end{tiny} & \begin{tiny}CFMc\end{tiny} & \begin{tiny}CFMe\end{tiny} & \begin{tiny}Cutoff\end{tiny} & \begin{tiny}UPL\end{tiny} & \begin{tiny}CFMc\end{tiny} & \begin{tiny}CFMe\end{tiny} & \begin{tiny}Cutoff\end{tiny} & \begin{tiny}UPL\end{tiny} & \begin{tiny}CFMc\end{tiny} & \begin{tiny}CFMe\end{tiny}\\\hline

\begin{scriptsize}\textbf{SOSP}\end{scriptsize} & \begin{scriptsize}$<10\%$\end{scriptsize} & $\checkmark$ & $\checkmark$ & $\checkmark$ & \begin{scriptsize}$<0.005$\end{scriptsize} & $\checkmark$ & $\checkmark$ & $\checkmark$ & \begin{scriptsize}$<1.73^\circ$\end{scriptsize} & $\boldsymbol{\times}$ & $\checkmark$ & $\checkmark$ & \begin{scriptsize}$<0.14$\end{scriptsize} & $\checkmark$ & $\boldsymbol{\times}$ & $\boldsymbol{\times}$ & \begin{scriptsize}$<7.19s$\end{scriptsize} & $\checkmark$ & $\checkmark$ & $\checkmark$\\\hline

\begin{scriptsize}\textbf{Head-On}\end{scriptsize} & \begin{scriptsize}$<10\%$\end{scriptsize} & $\checkmark$ & $\checkmark$ & $\checkmark$ & \begin{scriptsize}$<0.005$\end{scriptsize} & $\checkmark$ & $\checkmark$ & $\checkmark$ & \begin{scriptsize}$<0.91^\circ$\end{scriptsize} & $\boldsymbol{\times}$ & $\checkmark$ & $\boldsymbol{\times}$ & \begin{scriptsize}$<0.05$\end{scriptsize} & $\checkmark$ & $\boldsymbol{\times}$ & $\boldsymbol{\times}$ & \begin{scriptsize}$<4.84s$\end{scriptsize} & $\checkmark$ & $\boldsymbol{\times}$ & $\boldsymbol{\times}$ \\\hline

\begin{scriptsize}\textbf{Parallel-Ped}\end{scriptsize} & \begin{scriptsize}$<10\%$\end{scriptsize} & $\checkmark$ & $\checkmark$ & $\checkmark$ & \begin{scriptsize}$<0.005$\end{scriptsize} & $\checkmark$ & $\checkmark$ & $\checkmark$ & \begin{scriptsize}$<1.31^\circ$\end{scriptsize} & $\boldsymbol{\times}$ & $\checkmark$ & $\checkmark$ & \begin{scriptsize}$<0.17$\end{scriptsize} & $\checkmark$ & $\boldsymbol{\times}$ & $\boldsymbol{\times}$ & \begin{scriptsize}$<9.41s$\end{scriptsize} & $\checkmark$ & $\checkmark$ & $\checkmark$ \\\hline

\begin{scriptsize}\textbf{MOSP}\end{scriptsize} & & & & & & & & & & & & & & & & & & & &\\

\begin{scriptsize}\textbf{Case A}\end{scriptsize} & \begin{scriptsize}$<10\%$\end{scriptsize} & $\checkmark$ & $\checkmark$ & $\checkmark$ & \begin{scriptsize}$<0.005$\end{scriptsize} & $\boldsymbol{\times}$ & $\checkmark$ & $\checkmark$ & \begin{scriptsize}$<1.29^\circ$\end{scriptsize} & $\boldsymbol{\times}$ & $\checkmark$ & $\checkmark$ & \begin{scriptsize}$<0.14$\end{scriptsize} & $\checkmark$ & $\boldsymbol{\times}$ & $\boldsymbol{\times}$ & \begin{scriptsize}$<5.81s$\end{scriptsize} & $\checkmark$ & $\boldsymbol{\times}$ & $\boldsymbol{\times}$ \\

\begin{scriptsize}\textbf{Case B}\end{scriptsize} & \begin{scriptsize}$<10\%$\end{scriptsize} & $\checkmark$ & $\checkmark$ & $\checkmark$ & \begin{scriptsize}$<0.005$\end{scriptsize} & $\boldsymbol{\times}$ & $\checkmark$ & $\checkmark$ & \begin{scriptsize}$<2.1^\circ$\end{scriptsize} & $\boldsymbol{\times}$ & $\boldsymbol{\times}$ & $\boldsymbol{\times}$ & \begin{scriptsize}$<0.35$\end{scriptsize} & $\checkmark$ & $\boldsymbol{\times}$ & $\boldsymbol{\times}$ & \begin{scriptsize}$<5.96s$\end{scriptsize} & $\checkmark$ & $\boldsymbol{\times}$ & $\boldsymbol{\times}$ \\

\begin{scriptsize}\textbf{Case C}\end{scriptsize} & \begin{scriptsize}$<10\%$\end{scriptsize} & $\checkmark$ & $\checkmark$ & $\boldsymbol{\times}$ & \begin{scriptsize}$<0.005$\end{scriptsize} & $\boldsymbol{\times}$ & $\checkmark$ & $\checkmark$ & \begin{scriptsize}$<2.61^\circ$\end{scriptsize} & $\boldsymbol{\times}$ & $\boldsymbol{\times}$ & $\boldsymbol{\times}$ & \begin{scriptsize}$<0.37$\end{scriptsize} & $\boldsymbol{\times}$ & $\boldsymbol{\times}$ & $\boldsymbol{\times}$ & \begin{scriptsize}$<6.74s$\end{scriptsize} & $\boldsymbol{\times}$ & $\boldsymbol{\times}$ & $\boldsymbol{\times}$ \\

\begin{scriptsize}\textbf{Case D}\end{scriptsize} & \begin{scriptsize}$<10\%$\end{scriptsize} & $\checkmark$ & $\boldsymbol{\times}$ & $\boldsymbol{\times}$ & \begin{scriptsize}$<0.005$\end{scriptsize} & $\boldsymbol{\times}$ & $\checkmark$ & $\checkmark$ & \begin{scriptsize}$<2.01^\circ$\end{scriptsize} & $\boldsymbol{\times}$ & $\boldsymbol{\times}$ & $\boldsymbol{\times}$ & \begin{scriptsize}$<0.24$\end{scriptsize} & $\boldsymbol{\times}$ & $\boldsymbol{\times}$ & $\boldsymbol{\times}$ & \begin{scriptsize}$<6.53s$\end{scriptsize} & $\boldsymbol{\times}$ & $\boldsymbol{\times}$ & $\boldsymbol{\times}$ \\\hline

\end{tabular}
\end{table*}

\begin{table}
\centering

\caption{{\footnotesize A model’s score for final evaluation is calculated by counting the number of $\checkmark$ it receives in Table \ref{table:stage2}. The evaluation process highlights that UPL shows the best results among the five force-based models, followed by CFMc and CFMe, while SFMc and SFMe are `Not Eligible (NE)' from Stage \Romannum{1}.}}
\label{table:scoring}

\scriptsize
\def\arraystretch{2}
\begin{tabular}{|c|c|c|c|}
\hline

Models & $\begin{large}\checkmark\end{large}$ & $\begin{large}\boldsymbol{\times}\end{large}$ & $\textbf{SCORE}$\\\hline

$\textbf{UPL}$ & \hspace{0.4cm}20\hspace{0.4cm} & \hspace{0.4cm}15\hspace{0.4cm} & $20/35 = 57.14\%$\\\hline

$\textbf{CFMc}$ & 19 & 16 & $19/35 = 54.29\%$\\\hline

$\textbf{CFMe}$ & 17 & 18 & $17/35 = 48.57\%$\\\hline

$\textbf{SFMc}$ & - & - & NE\\\hline

$\textbf{SFMe}$ & - & - & NE\\\hline

\end{tabular}
\end{table}

In Table \ref{table:stage1}, the first stage of the models' eligibility criteria is reported. For each experimental scenario, models achieving more than 80\% successful cases are marked as $\checkmark$; otherwise, they are marked as $\times$. Models with a number of $\checkmark$ greater than or equal to the number of $\times$ are shortlisted for further evaluation in Stage \Romannum{2}; Otherwise considered as `Not Eligible (NE)'. This eligibility criterion primarily determines if a model can provide simulation results comparable to the experimental results for the situations considered. Table \ref{table:stage1} clearly indicates that the UPL, CFMc, and CFMe models perform well ($>80\%$ successful trajectories) for most situations considered. Note that this study does not imply that the remaining models are entirely ineffective, but rather that they are not suitable for these common situations with their current parametric values (refer to Section C in the Supplementary material for more details). A search across the parameter space of the models is beyond the scope of this article. Interestingly, almost all models are unsuccessful for the MOSP scenarios with higher obstacle densities.

In the second stage, $\checkmark$ and $\times$ marks are again assigned to the shortlisted models, based on whether their average metric values fall within the specified cutoff values (refer to Table \ref{table:stage2}). Cutoff values for path smoothness, speed deviation, and travel time are determined by using the $3\sigma$ rule, adding three standard deviations to the mean values observed during the experiments, while the cutoff values for overlap and oscillation metrics are set at 10\% and 0.005, respectively, due to no observed values for these metrics during experiments.

Normalized success score for a model is derived by counting the number of $\checkmark$ it receives (refer to Table \ref{table:scoring}).  The results after the two-stage process show that UPL yields the best outcomes among the five force-based models, followed by CFMc and CFMe. However, the final scores of UPL, CFMc and CFMe are separated by $<10\%$, suggesting that they yield comparable predictions for the situations considered here. It is important to recognize the significance of the first stage of shortlisting. If scoring had been directly applied to all models, SFMc would have received the highest score despite failing to achieve 80\% successful trajectories for all the setups, including the simplest SOSP scenario. However, our analysis also indicates the failure of all models for the higher-density MOSP scenarios.

\section{Conclusions}

Thus in this study, we have provided a comprehensive scoring system designed to evaluate pedestrian dynamics models for commonly occurring situations involving moderate-to-low pedestrian density (with large obstacle density for some cases). Towards this, we have selected four common situations and performed experiments with volunteers spanning age groups and gender. The experimental results serve as a baseline to develop a two-stage evaluation process including eligibility criteria and scoring. We provide a detailed evaluation of five force-based models by mimicking experimental scenarios using exact initial and final conditions. Our overall observations from the experiments are summarized below.

\begin{itemize}
    \item SOSP: SFMc fails to produce 80\% successful trajectories and shows high overlap. SFMe passes the eligibility but demonstrates abrupt path changes, significant speed deviations, and longer travel times. UPL results in sudden path changes but minimal speed deviation, while CFMc and CFMe exhibit high speed deviations but low path direction changes. Nearly all models show a significant decrease in speed as the pedestrian approaches the obstacle.

    \item Head-On: Both SFMc and SFMe fail to achieve 80\% success rates. CFMe shows some unsuccessful trajectories, whereas UPL and CFMc have none. CFMe also shows abrupt path changes, significant speed deviations, and increased travel times. UPL resulted in sudden path change but minimal speed deviation, whereas CFMc shows high speed deviation but low change in path direction.

    \item Parallel-Ped: SFMc fails to reach an 80\% success rate and has high overlap, while all other models show 100\% success. UPL causes sudden path changes with minimal speed deviations, whereas the remaining models display high speed deviations with low path direction changes. In experiments, the faster pedestrian overtakes the slower one, without any significant change in the dynamics of the slower one. For all models except UPL, the faster pedestrian couldn’t overtake the slower one. UPL predicts an overtake but with a significant path change of B which is not observed in experiments.

    \item MOSP: With increasing obstacle density, all models showed higher proportions of unsuccessful trajectories, greater path complexity, speed deviations, and longer travel times, although these effects depend on specific obstacle configurations and pedestrian positions. Both SFMc and SFMe fail to achieve 80\% success in nearly all cases, while other models perform relatively better. UPL exhibits backward motion and abrupt path changes, whereas CFMc and CFMe have sudden path changes and high speed deviations.
\end{itemize}

Here, we note that nearly all models fail for MOSP situations with a large number of obstacles (14.96\%). This is a highly common scenario in India, where one needs to navigate through moderately crowded public spaces (even simple corridors with a moderate density of standing people). While it is true that parameter tuning (e.g., adjusting k and $\tau_0$ in the case of UPL) can influence model behavior, our additional analysis (see Figure E-8 in Supplementary material) demonstrates that even significant variations (up to tenfold changes) in these parameters do not substantially improve the model's performance in the tested scenarios. This supports the article’s focus on addressing deeper structural issues in the models, rather than parameter sensitivity alone.

Next, we finalize our evaluation by a two-stage process explained in Tables \ref{table:stage1}, \ref{table:stage2}, and \ref{table:scoring}. Overall, the UPL \cite{karamouzas2014universal} emerged as the most successful in our scoring system, which might provide decent predictions in the most common situations, closely followed by CFMc and CFMe.

Note that, while we state that UPL performs well relative to other models, notable limitations are visible within this study. In particular, UPL consistently demonstrates sudden changes in the pedestrian path across all scenarios, which indicates the unrealistic feature for collision avoidance. In SOSP, UPL predicts a slowdown of the pedestrian as it approaches the obstacle. For Parallel-Ped scenarios, UPL predicts a path change of B (and subsequent overtake of A), which contrasts experiments. Furthermore, in the multiple obstacle scenario (MOSP), UPL exhibits noticeable oscillations, suggesting an unrealistic fluctuation in pedestrian movement when navigating around obstacles. Additionally, UPL predicts increasingly unsuccessful trajectories for higher obstacle densities in MOSP situations. These features imply difficulties in accurately replicating the decision-making process and adaptive behaviors observed in real-life pedestrian dynamics. Thus even though UPL provides the most reasonable predictions in most scenarios, it shows significant limitations that need to be corrected. To the best of our knowledge, no other study has compared all these models for such commonly occurring situations. Thus, this study will help future researchers in the selection of appropriate pedestrian models for a given scenario.

\section{Supplementary Material}

The supplementary material offers further insights to enhance the findings discussed in the main manuscript. While the main text sets a benchmark for pedestrian dynamics models in common scenarios, the supplementary section addresses several key aspects: the determination of the time step size used in simulations, the unrealistic outcomes produced by models in the MOSP scenario, parameter selection for the Social Force Model (SFM), and additional trajectory data for the SOSP scenario.

\section{Acknowledgement}

The authors would like to extend their appreciation to all the students of IIT Kanpur who participated as volunteers in this study. We are also deeply grateful to the security personnel and administration of IIT Kanpur for their invaluable assistance in conducting the experiments. Additionally, we acknowledge the foundational contributions made by former dual-degree students of IIT Kanpur—Ishan Prashant, Amullya Kale, and Satyendra Pandey—whose work significantly informed this study. This research was supported by the CRG project (Sanction number: IME/SERB/2023137) from the Science and Engineering Research Board (SERB), Department of Science and Technology (DST), India.\\\\
$^*$Email IDs to the corresponding authors are sprawesh@iitk.ac.in, indrasd@iitk.ac.in, and anuragt@iitk.ac.in.

\bibliographystyle{ieeetr}
\bibliography{references}

\end{document}


\title{Appendix}

\date{\vspace{-5ex}}

\maketitle

\section{Determination of timestep size}

\begin{figure} [h]
    \centering
    \includegraphics[width= 1\linewidth]{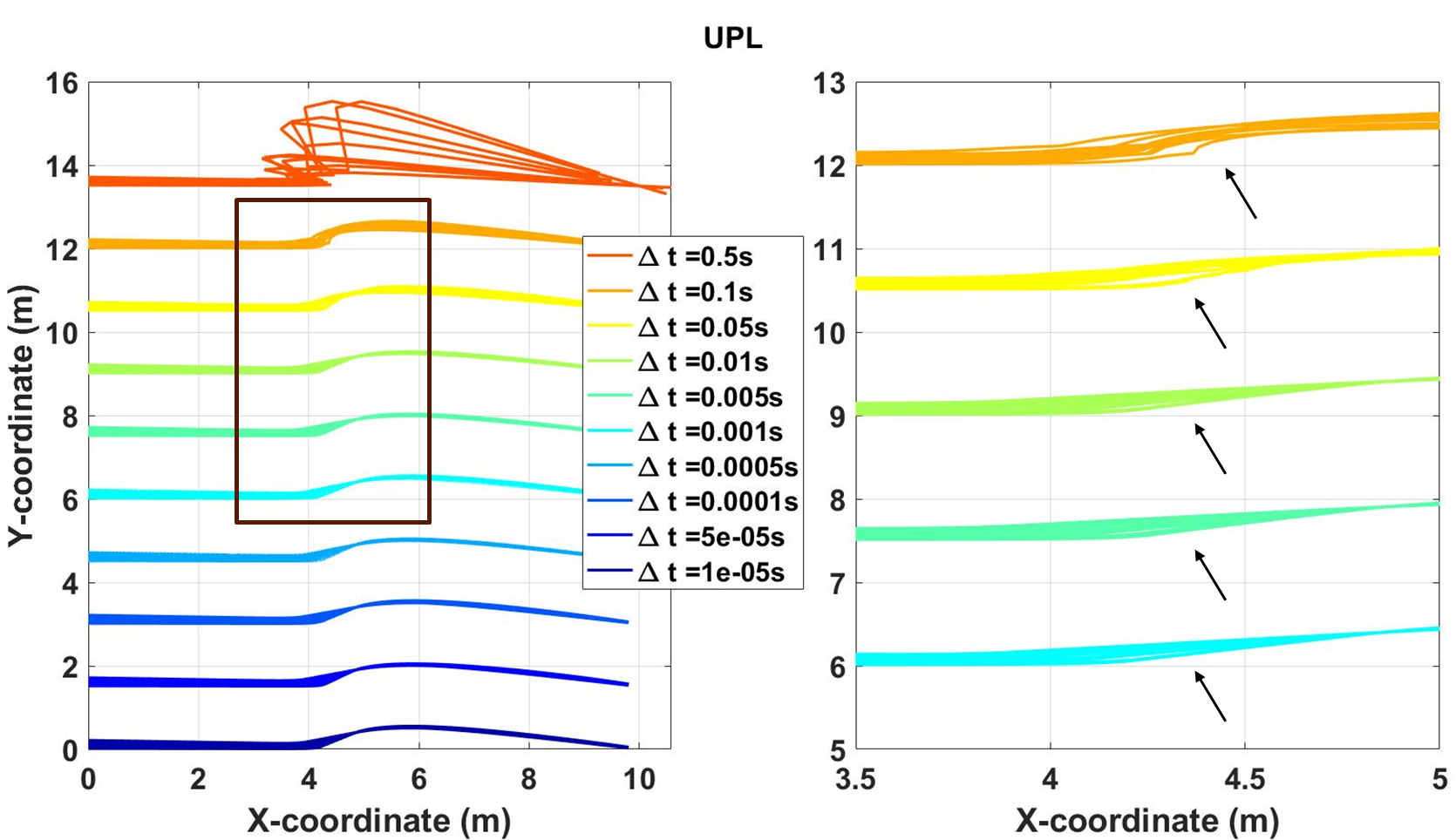}
        
    \caption{{\footnotesize Trajectories of SOSP case with different values of $\Delta t$ simulated using UPL model (right). A magnified view of trajectories within the black rectangular box is given on the left side, highlighting the convergence of trajectories after $\Delta t = 0.01\;s$}} \label{fig:delt_UPL}
    
\end{figure}

\begin{figure} [h]
    \centering
    \includegraphics[width= 1\linewidth]{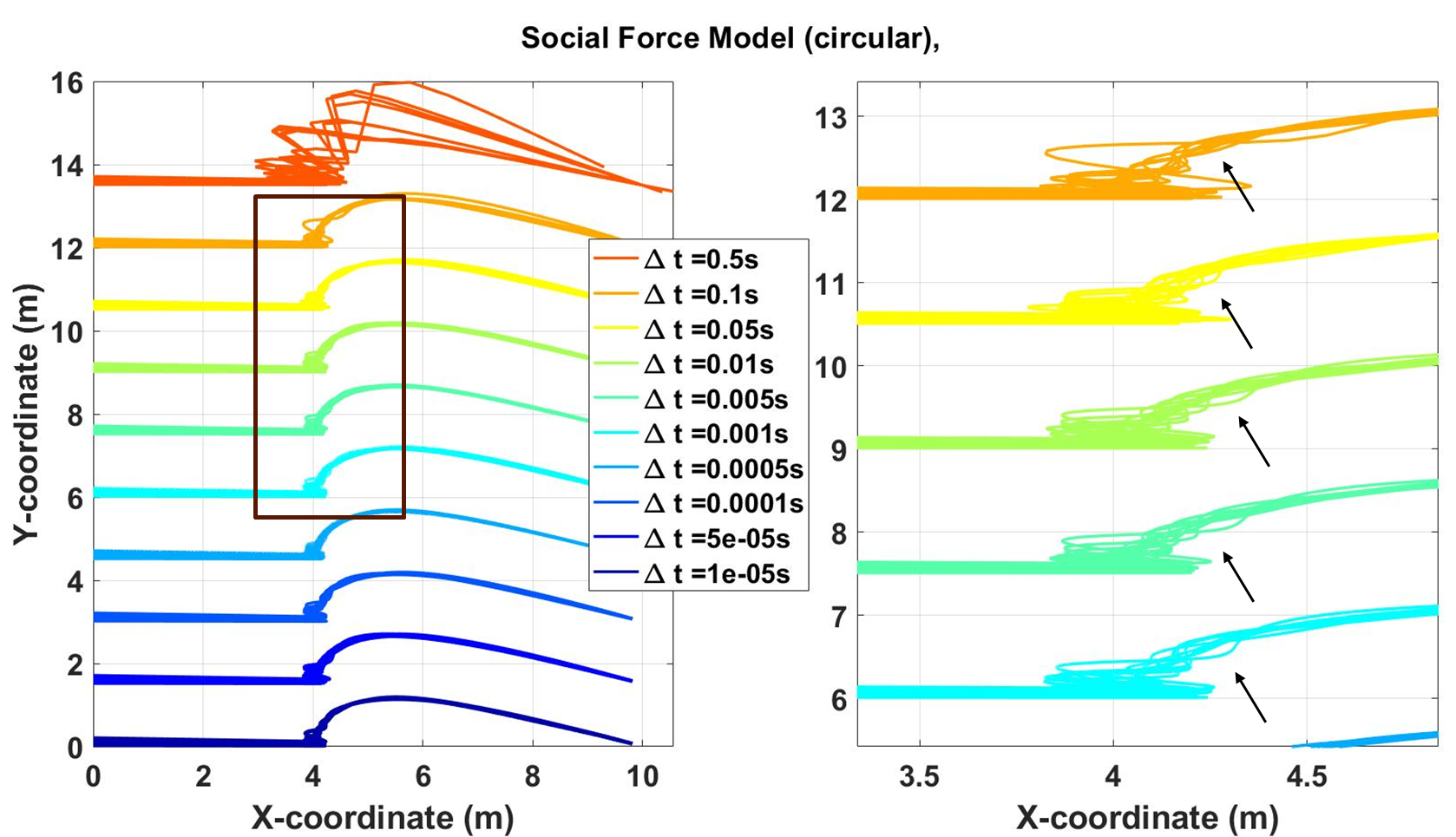}
        
    \caption{{\footnotesize Trajectories of SOSP case with different values of $\Delta t$ simulated using SFMc model (right). A magnified view of trajectories within the black rectangular box is given on the left side, highlighting the convergence of trajectories after $\Delta t = 0.01\;s$}} \label{fig:delt_SFMc}
    
\end{figure}

\begin{figure} [h!]
    \centering
    \includegraphics[width= 1\linewidth]{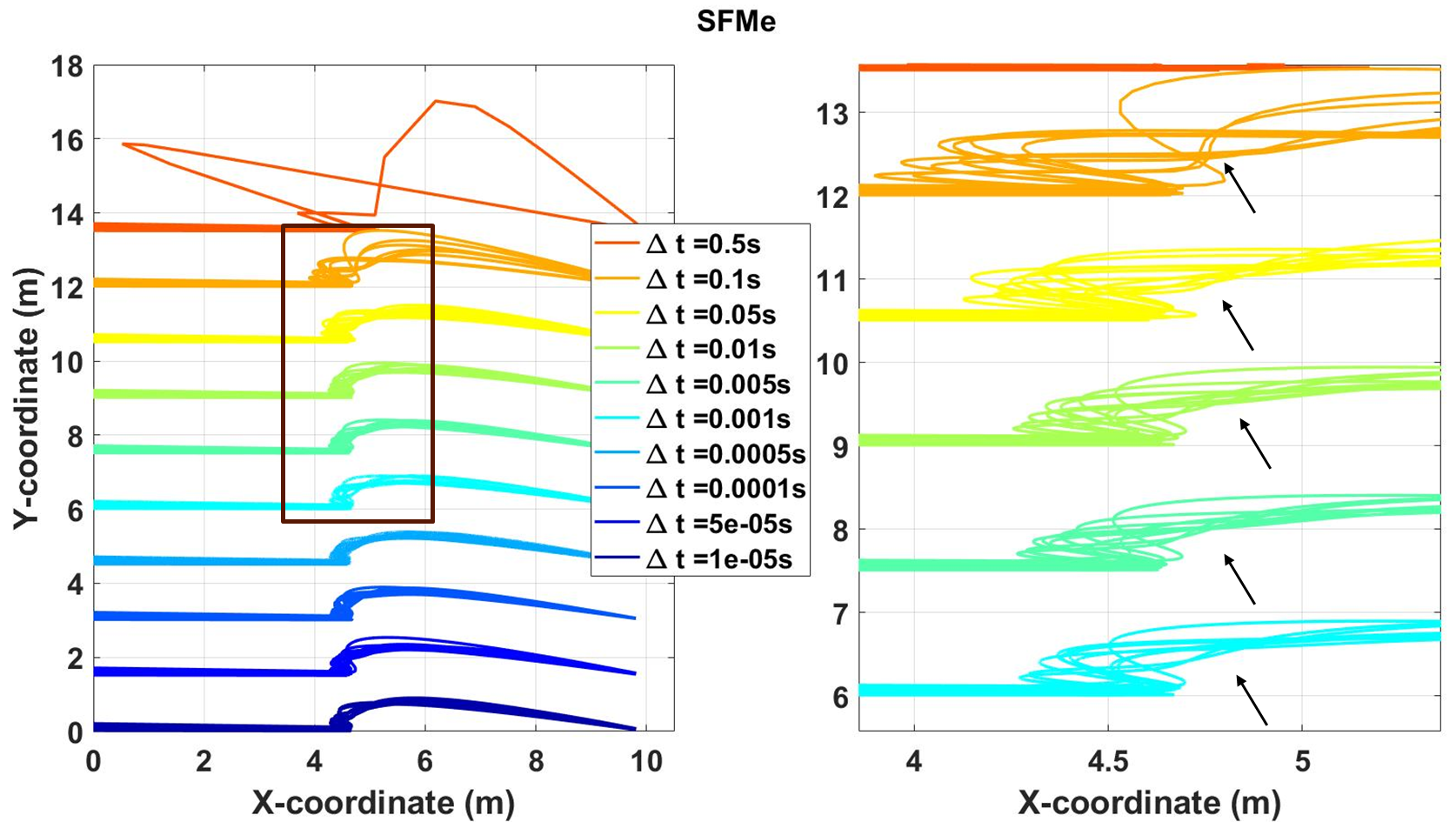}
        
    \caption{{\footnotesize Trajectories of SOSP case with different values of $\Delta t$ simulated using SFMe model (right). A magnified view of trajectories within the black rectangular box is given on the left side, highlighting the convergence of trajectories after $\Delta t = 0.01\;s$}} \label{fig:delt_SFMe}
    
\end{figure}

In this section, the value of $\Delta t$ for UPL, SFMc, and SFMe is estimated in order to avoid numerical instability in such simulations. Multiple simulations are performed by varying $\Delta t$ from 0.5 s to $10^{-5}$ s (Figures \ref{fig:delt_UPL}-\ref{fig:delt_SFMe}). Note that ten different offsets (deviation from the original position) are simulated for each $\Delta t$ since the behavior strongly depends on the magnitude of the offset. If we look at the figures, we will observe that in the cases of UPL, SFMc, and SFMe, the trajectories converge at 0.01s. Zoomed figures are also plotted in figures. The author has also set the value of time iteration as 0.01s in the cases of CFMc and CFMe. Hence, $\Delta t =  0 .01s$ is set for all the models in this study.

\newpage

\section{Simulation Results for MOSP}

\begin{figure} [h]
    \centering
    \includegraphics[width= 1\linewidth]{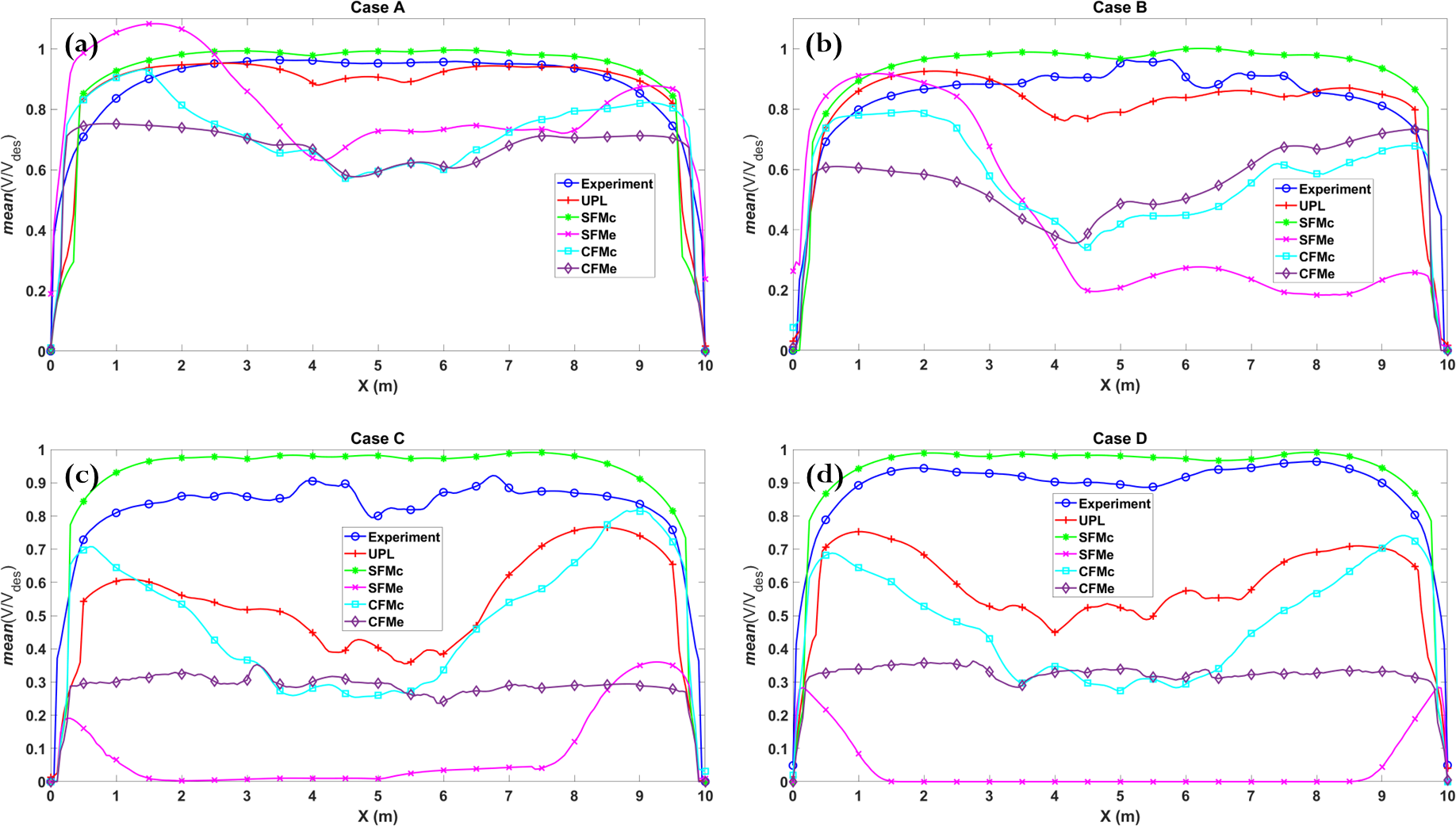}
        
    \caption{{\footnotesize Normalized speed averaged over all the trajectories are plotted with respect to x coordinates of the pedestrian position for all four cases.}} \label{fig:MOSP_avgVel}
    
\end{figure}

In Figure \ref{fig:MOSP_avgVel}, normalized speed averaged over all the trajectories are plotted with respect to x coordinates of the pedestrian position for all four cases. During the experiment, pedestrians tend to maintain their desired speed, however, minor irregularities and speed deviations can be observed due to frequent changes in direction caused by the presence of multiple obstacles. In contrast, all the models exhibit significant deviations from the desired speed, especially in high-density cases. Notably, the SFMc model displays a near-perfect speed profile, which is attributed to its weak interaction force, allowing pedestrians to pass through the obstacle maze by overlapping with obstacles without altering their path, also shown by a trajectory in Figure \ref{fig:SFMcSFMe_Traj_MOSP}(a). Additionally, the SFMe model shows an average speed approaching $0\;m/s$ in Cases C and D due to excessively large interaction forces that immobilize pedestrians even before they enter the obstacle maze, also observed by the pedestrian trajectory in Figure \ref{fig:SFMcSFMe_Traj_MOSP}(b).

\subsection*{Social Force Model unrealistic behaviour}

\begin{figure} [h]
    \centering
    \subfigure[]
    {
        \includegraphics[width= 0.8\linewidth]{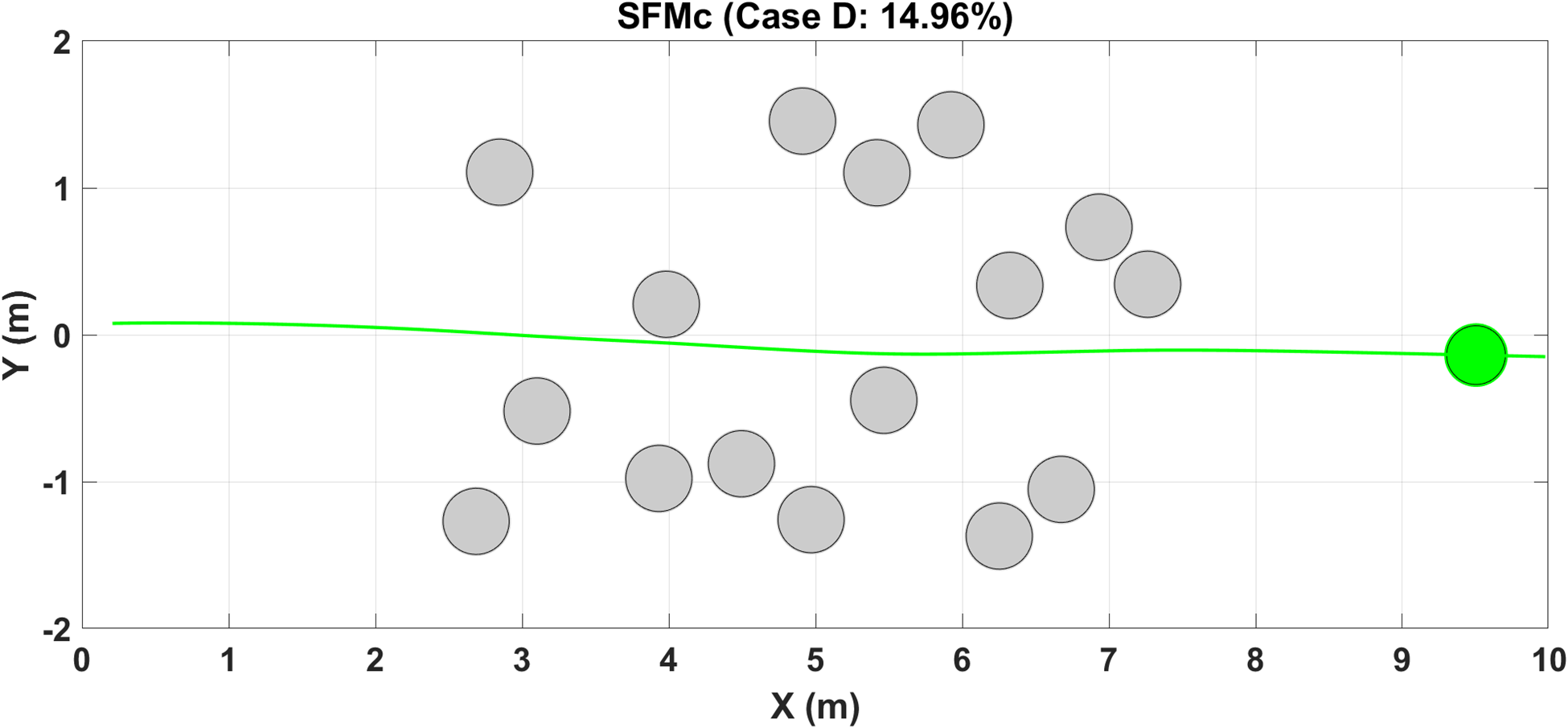}
    }

    \subfigure[]
    {
        \includegraphics[width= 0.8\linewidth]{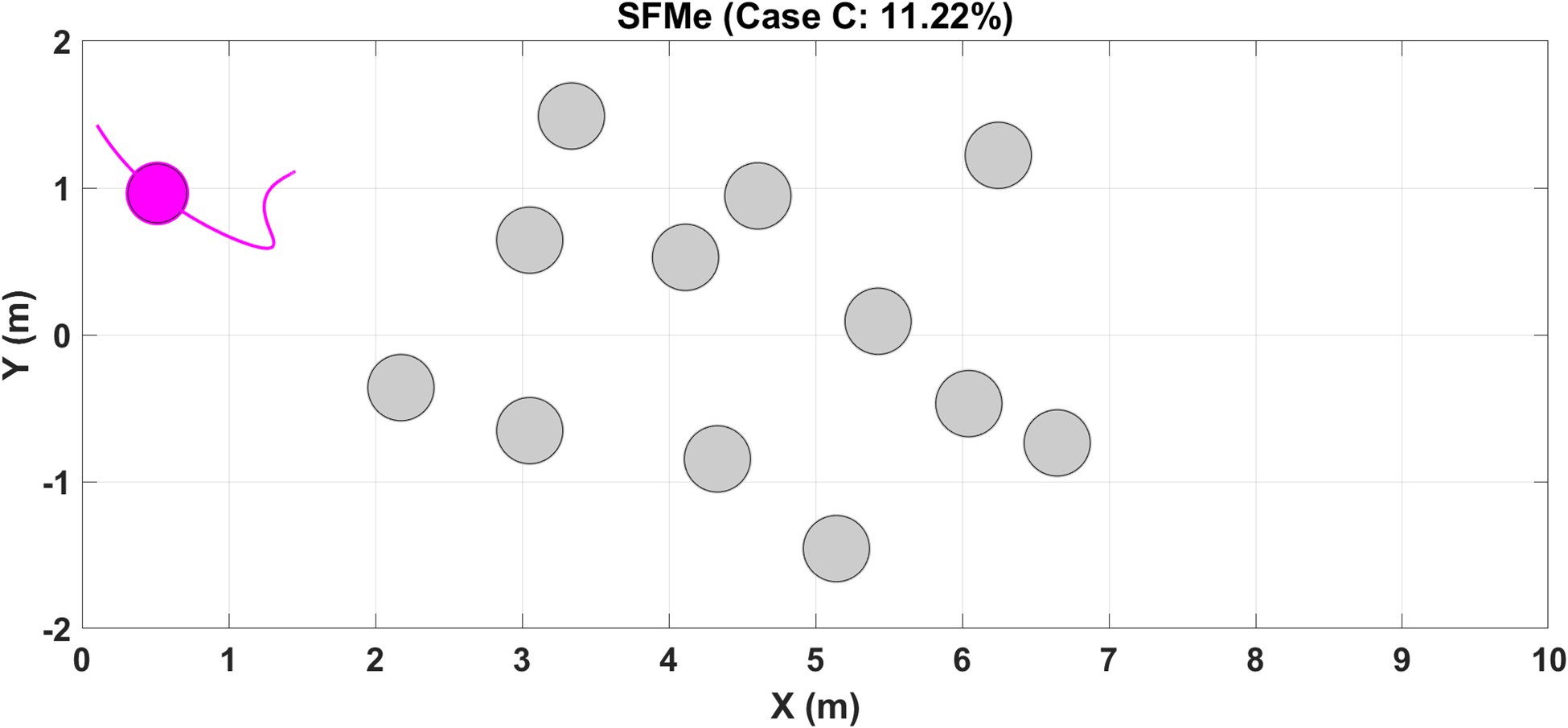}
    }

    \caption{Trajectories graphs for a) SFMc (MOSP Case D) and b) SFMe (MOSP Case C), showing unrealistic simulation results.}
    \label{fig:SFMcSFMe_Traj_MOSP}
    
\end{figure}

During simulations, both the SFMc and SFMe models exhibited unrealistic results, particularly in MOSP scenarios. In Figure \ref{fig:SFMcSFMe_Traj_MOSP}(a), one of the trajectories of the SFMc model for MOSP Case D is shown. This trajectory indicates that the pedestrian passes through the maze of obstacles without altering its path to avoid them, resulting in overlaps. Consequently, as illustrated in the figure, the SFMc model only registers successful cases when the obstacle configuration is favorable, rather than attempting to avoid obstacles. Similarly, Figure \ref{fig:SFMcSFMe_Traj_MOSP}(b), shows one of the trajectories for the SFMe model for MOSP Case C. This trajectory shows the pedestrian becoming immobilized without even entering the measurement zone. For the SFMe model, the interaction forces' values become excessively large due to the presence of multiple obstacles, rendering the pedestrian stationary, which is clearly unrealistic.

\newpage

\section{Parameter selection for Social Force Models}

\begin{figure} [h]
    \includegraphics[width= 1\linewidth]{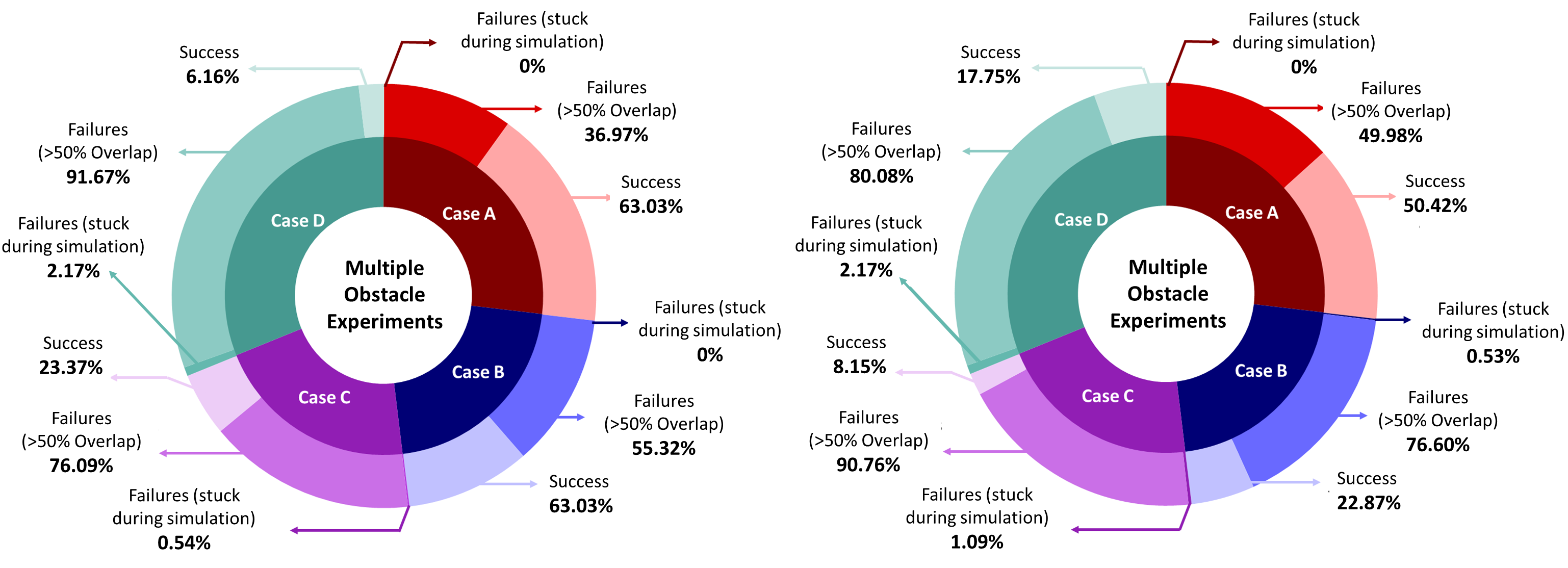}    \caption{{\footnotesize Detailed pie charts for (a) SFMc and (b) SFMe when angular dependency is considered during simulation. Multiple obstacle experiments are reproduced using SFMc and SFMe for four different cases: Case A (3.74\% of area covered by obstacles), Case B (6.54\% of area), Case C (11.22\% of area), and Case D (14.96\% of area). The figures show a large share of failed simulations in all four cases, with most failures due to more than 50\% overlap with obstacles, while only a few (or none in some cases) are due to simulations getting stuck. This indicates that pedestrians are unable to effectively avoid the obstacles, resulting in significant overlap when utilizing the social force model with angular dependency.}} \label{fig:SFMcSFMe_lamda}
\end{figure}

The values of parameters for SFMc and SFMe were taken from the study by Johansson et. al. \cite{johansson2007specification}. The authors provided two sets of calibrated values for the constants A and B. One set assumes no angular dependency in the social force ($\lambda = 1$), while the other set includes angular dependency ($\lambda \neq 1$). Ideally, angular dependency should be considered in any social force model to accurately represent pedestrian behavior. However, our simulations of multiple obstacle scenarios using SFMc and SFMe with $\lambda \neq 1$ revealed significant issues. Detailed pie charts illustrating findings for SFMc and SFMe are shown in Figure \ref{fig:SFMcSFMe_lamda}.  We observe a large share of failed simulations with a low incidence of failures due to pedestrians getting stuck. Instead, most failures occurred because trajectories resulted in $\geq 50\%$ overlap with obstacles. This suggests that the magnitude of the forces experienced by pedestrians was insufficient to avoid obstacles, leading to excessive overlap. This outcome indicates a fundamental problem in social force failing to help avoid the obstacles when angular dependency is considered. Consequently, for our study, we opted to use the calibrated values of constants A and B for the case where $\lambda = 1$.

\newpage

\section{Single Pedestrian Single Obstacle}

\begin{figure}
    \centering
    \subfigure[Universal Power Law (UPL)]
    {
        \includegraphics[width= 1\linewidth]{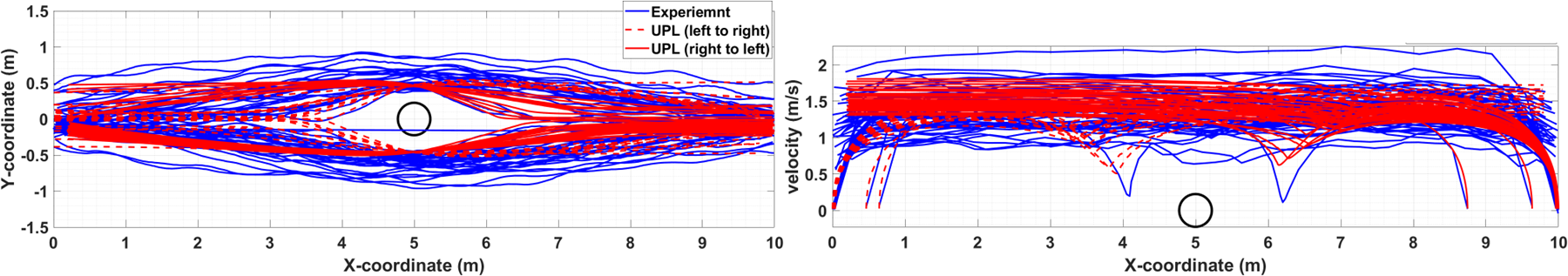}
        \label{fig:singleObst_reg_UPL}
    }

    \subfigure[Social Force Model circular (SFMc)]
    {
        \includegraphics[width= 1\linewidth]{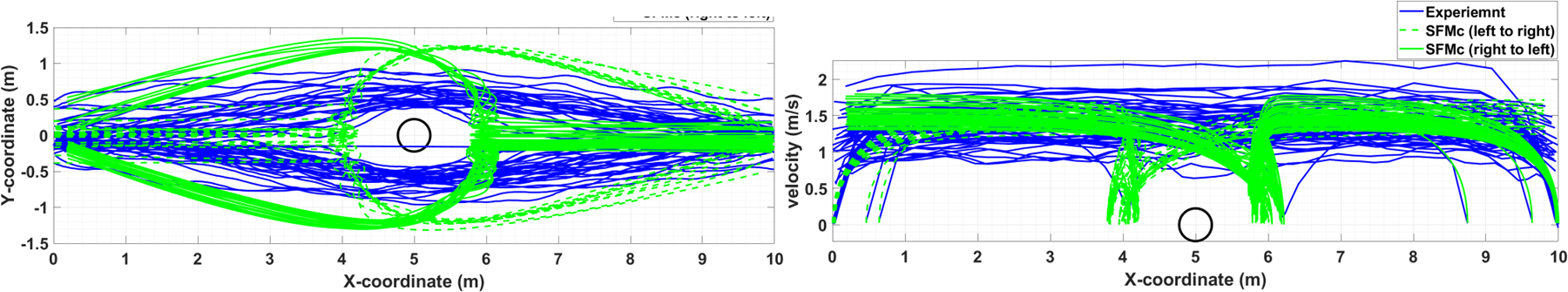}
        \label{fig:singleObst_reg_SFMc}
    }

    \subfigure[Social Force Model elliptical (SFMe)]
    {
        \includegraphics[width= 1\linewidth]{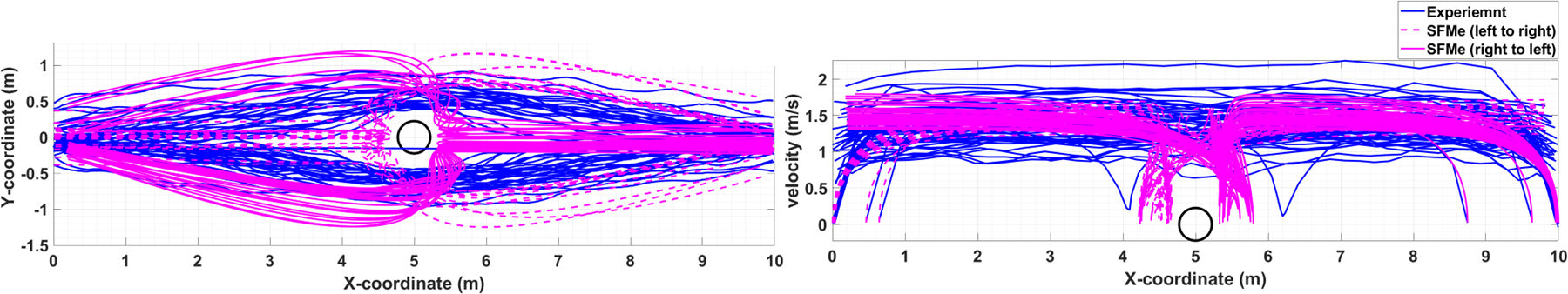}
        \label{fig:singleObst_reg_SFMe}
    }
    \subfigure[Centrifugal Force Model circular (CFMc)]
    {
        \includegraphics[width= 1\linewidth]{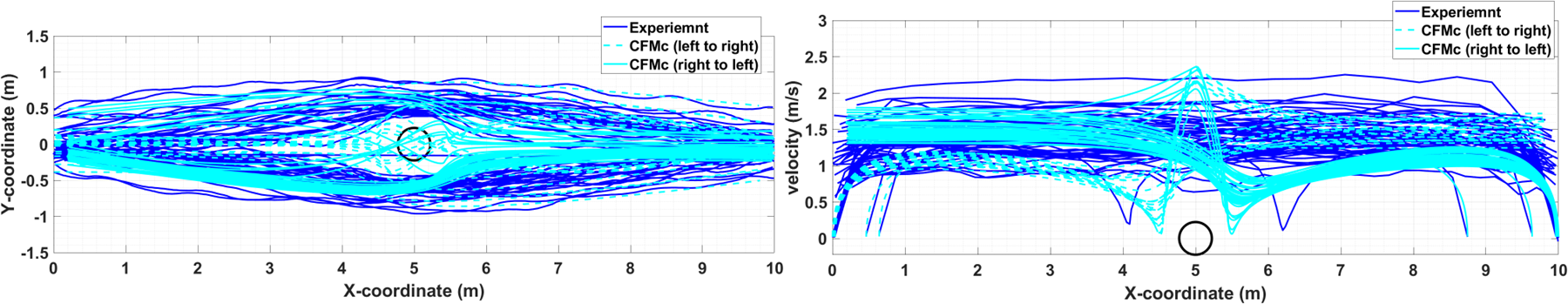}
        \label{fig:singleObst_reg_CFMc}
    }
    \subfigure[Centrifugal Force Model elliptical (CFMe)]
    {
        \includegraphics[width= 1\linewidth]{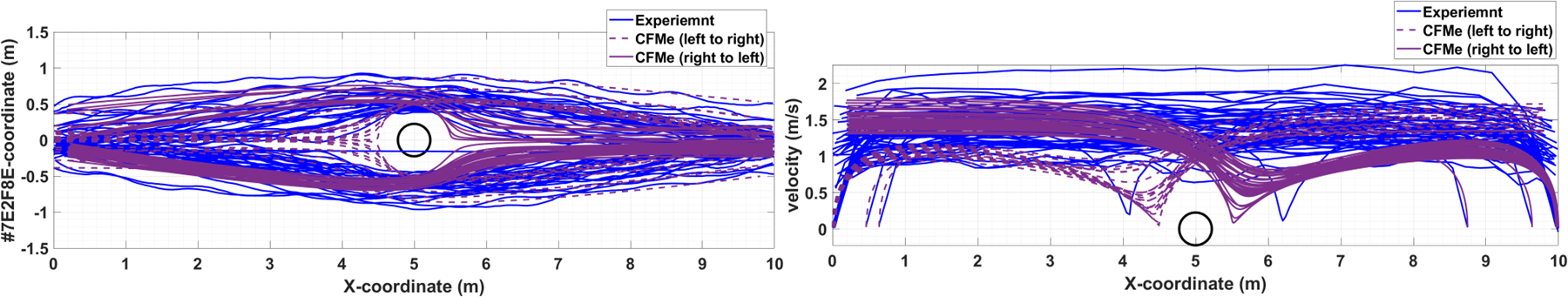}
        \label{fig:singleObst_reg_CFMe}
    }
    \caption{Trajectories and velocity graphs are generated using all the force-based models and are compared with the experiment.} \label{fig:singleObst_reg}
    
\end{figure}

All the trajectories and velocities generated by the reproducing experiment are shown in Figure \ref{fig:singleObst_reg}. In the case of UPL, neither oscillations nor overlaps are present in any of the trajectories. However, the sudden dip in speeds while crossing the obstacle can be observed, especially in the cases of low offset (Figure \ref{fig:singleObst_reg}(a)). For SFMc, we can see the sudden change in direction and multiple oscillations in each and every trajectory (Figure \ref{fig:singleObst_reg}(b)). Similarly, in the case of SFMe, sudden changes in directions and oscillations are present. However, in this case, pedestrians reach closer to the obstacle, and the number of times pedestrians oscillate is lower, (Figure \ref{fig:singleObst_reg}(c)). No oscillations are predicted in the case of centrifugal force models, but overlaps are present, especially for the low offsets in the case of CFMc (Figure \ref{fig:singleObst_reg}(d)) and sudden change in directions in the case of CFMe, (Figure \ref{fig:singleObst_reg}(c)).

\section{Model evaluation for different parameters}

\begin{figure} [h]
    \includegraphics[width= 1\linewidth]{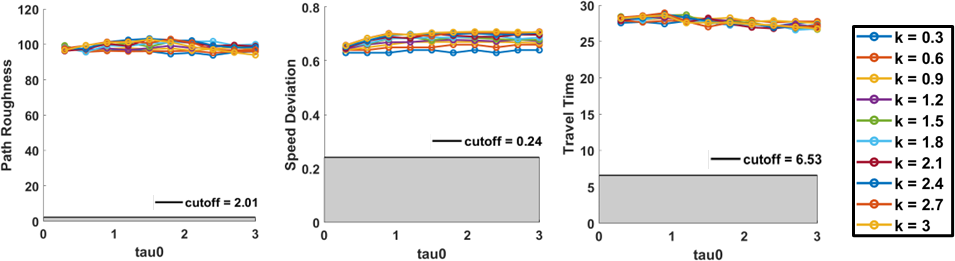}    \caption{{\footnotesize The values of evaluating metrics by varying parameters k and tau0 in the case of UPL for the MOSP (Case D) scenario.}} \label{fig:UPL_evaluation}
\end{figure}

While it is true that parameter tuning (e.g., adjusting k and tau0 in the case of UPL) can influence model behavior, our additional analysis (see Figure \ref{fig:UPL_evaluation}) demonstrates that even significant variations (up to tenfold changes) in these parameters do not substantially improve the model's performance in the tested scenarios. This supports the article’s focus on addressing deeper structural issues in the models, rather than parameter sensitivity alone.

\pagebreak

\bibliographystyle{plain}
\bibliography{references}